\documentclass[twocolumn]{aastex631}
\usepackage{amsmath}
\usepackage{multirow}
\usepackage{rotating}

\begin{document}

\title{Constraining possible $\gamma$--ray burst emission from GW230529 using \textit{Swift}--BAT and \textit{Fermi}--GBM}

\author[0000-0003-0020-687X]{Samuele Ronchini}
\affiliation{Department of Astronomy and Astrophysics, 
The Pennsylvania State University, 525 Davey Lab, University Park, PA 16802, USA} 
\affiliation{Institute for Gravitation \& the Cosmos, The Pennsylvania State University, University Park PA 16802, USA}

\author[0000-0002-6657-9022]{Suman Bala}
\affiliation{Science and Technology Institute, Universities Space Research Association, Huntsville, AL 35805, USA}

\author[0000-0001-9012-2463]{Joshua Wood}
\affiliation{NASA Marshall Space Flight Center, Huntsville, Al 35812, USA}

\author[0000-0001-5229-1995]{James Delaunay}
\affiliation{Department of Astronomy and Astrophysics, 
The Pennsylvania State University, 525 Davey Lab, University Park, PA 16802, USA}

\author[0000-0001-6849-1270]{Simone Dichiara}
\affiliation{Department of Astronomy and Astrophysics, 
The Pennsylvania State University, 525 Davey Lab, University Park, PA 16802, USA}

\author[0000-0002-6745-4790]{Jamie A. Kennea}
\affiliation{Department of Astronomy and Astrophysics, 
The Pennsylvania State University, 525 Davey Lab, University Park, PA 16802, USA}

\author[0000-0002-4299-2517]{Tyler Parsotan}
\affiliation{Astrophysics Science Division, NASA Goddard Space Flight Center, Greenbelt, MD 20771, USA}

\author[0000-0003-0852-3685]{Gayathri Raman}
\affiliation{Department of Astronomy and Astrophysics, 
The Pennsylvania State University, 525 Davey Lab, University Park, PA 16802, USA}

\author[0000-0002-2810-8764]{Aaron Tohuvavohu}
\affiliation{Department of Astronomy \& Astrophysics, University of Toronto, Toronto, ON M5S 3H4}
\affiliation{Dunlap Institute for Astronomy \& Astrophysics, University of Toronto, Toronto, ON M5S 3H4}

\author[0000-0002-4559-8427]{Naresh Adhikari} 
\affiliation{Leonard E. Parker Center for Gravitation, Cosmology, and Astrophysics, University of Wisconsin-Milwaukee, Milwaukee, WI 53201, USA}

\author[0000-0001-7916-2923]{Narayana~P.~Bhat} 
\affiliation{Center for Space Plasma and Aeronomic Research, University of Alabama in Huntsville, Huntsville, AL 35899, USA}

\author[0000-0001-7616-7366]{Sylvia Biscoveanu}\affiliation{Center for Interdisciplinary Exploration and Research in Astrophysics (CIERA), Northwestern University, Evanston, IL 60201, USA}

\author[0000-0001-9935-8106]{Elisabetta Bissaldi}
\affiliation{Dipartimento Interateneo di Fisica, Politecnico di Bari, Via G. Amendola 173, 70125 Bari, Italy}
\affiliation{Istituto Nazionale di Fisica Nucleare, Sezione di Bari, Via E. Orabona 4, 70125 Bari, Italy}

\author[0000-0002-2942-3379]{Eric~Burns} 
\affil{Department of Physics \& Astronomy, Louisiana State University, Baton Rouge, LA 70803, USA}

\author[0000-0001-6278-1576]{Sergio Campana}
\affiliation{INAF-Osservatorio Astronomico di Brera, Via E. Bianchi 46, 23807 Merate, LC, Italy}

\author[0000-0003-4750-5551]{Koustav Chandra}
\affiliation{Institute for Gravitation \& the Cosmos, The Pennsylvania State University, University Park PA 16802, USA}

\author[0009-0003-3480-8251]{William H. Cleveland}
\affiliation{Science and Technology Institute, Universities Space Research Association, Huntsville, AL 35805, USA}

\author[0000-0003-1835-570X]{Sarah~Dalessi}
\affiliation{Department of Space Science, University of Alabama in Huntsville, 320 Sparkman Drive, Huntsville, AL 35899, USA}
\affiliation{Center for Space Plasma and Aeronomic Research, The University of Alabama in Huntsville, Huntsville, AL 35899}

\author{Massimiliano De Pasquale}
\affiliation{MIFT Department, Polo Papardo, University of Messina, Viale Ferdinando Stagno d'Alcontres, 31, 98166 Messina, Italy}

\author[0000-0002-9370-8360]{Juan Garc\'ia-Bellido}
\affiliation{Instituto de F\'isica Te\'orica UAM/CSIC, Universidad Aut\'onoma de Madrid, Cantoblanco 28049 Madrid, Spain}

\author[0000-0001-8335-9614]{Claudio Gasbarra}
\affiliation{Dipartimento di Fisica, Università di Roma Tor Vergata, I-00133 Roma, Italy}
\affiliation{Istituto Nazionale di Fisica Nucleare, Sezione di Roma Tor Vergata, I-00133 Roma, Italy}

\author{Misty M. Giles} 
\affiliation{Jacobs Space Exploration Group, Huntsville, AL 35806, USA}

\author[0000-0001-6932-8715]{Ish Gupta}
\affiliation{Institute for Gravitation \& the Cosmos, The Pennsylvania State University, University Park PA 16802, USA}

\author[0000-0002-8028-0991]{Dieter Hartmann}
\affiliation{Department of Physics \& Astronomy, Clemson University, Kinard Lab of Physics, Clemson, SC 29634, USA}

\author[0000-0001-9556-7576]{Boyan~A.~Hristov}
\affiliation{Center for Space Plasma and Aeronomic Research, The University of Alabama in Huntsville, Huntsville, AL 35899}

\author[0000-0002-0468-6025]{Michelle~C.~Hui}
\affiliation{NASA Marshall Space Flight Center, Huntsville, Al 35812, USA}

\author[0000-0002-5700-282X]{Rahul Kashyap}
\affiliation{Institute for Gravitation \& the Cosmos, The Pennsylvania State University, University Park PA 16802, USA}

\author[0000-0001-9201-4706]{Daniel Kocevski}
\affiliation{NASA Marshall Space Flight Center, Huntsville, Al 35812, USA}

\author[0000-0002-2531-3703]{Bagrat Mailyan}
\affiliation{Department of Aerospace, Physics and Space Sciences, Florida Institute of Technology, Melbourne, FL 32901, USA}

\author{Christian~Malacaria }
\affiliation{International Space Science Institute, Hallerstrasse 6, 3012 Bern, Switzerland }

\author[0000-0001-7665-0796]{Hiroyuki Nakano}
\affiliation{Faculty of Law, Ryukoku University, Kyoto 612-8577, Japan}

\author[0000-0003-0406-7387]{Giacomo Principe}
\affiliation{Dipartimento di Fisica, Universit\'a di Trieste, I-34127 Trieste, Italy}
\affiliation{Istituto Nazionale di Fisica Nucleare, Sezione di Trieste, I-34127 Trieste, Italy}
\affiliation{INAF - Istituto di Radioastronomia, I-40129 Bologna, Italy}

\author[0000-0002-7150-9061]{Oliver J. Roberts}
\affiliation{Science and Technology Institute, Universities Space and Research Association, 320 Sparkman Drive, Huntsville, AL 35805, USA.}

\author[0000-0003-3845-7586]{Bangalore Sathyaprakash}
\affiliation{Institute for Gravitation \& the Cosmos, The Pennsylvania State University, University Park PA 16802, USA}

\author[0000-0002-1334-8853]{Lijing Shao}
\affiliation{Kavli Institute for Astronomy and Astrophysics, Peking University, Beijing 100871, China} 
\affiliation{National Astronomical Observatories, Chinese Academy of Sciences, Beijing 100012, China}

\author[0000-0002-1869-7817]{Eleonora Troja}
\affiliation{Department of Physics, University of Rome Tor Vergata, 00100 Rome, Italy}
\affiliation{Istituto Nazionale di Astrofisica, 00100 Rome, Italy}

\author[0000-0002-2149-9846]{P\'eter~Veres}
\affiliation{Department of Space Science, University of Alabama in Huntsville, Huntsville, AL 35899, USA}
\affiliation{Center for Space Plasma and Aeronomic Research, University of Alabama in Huntsville, Huntsville, AL 35899, USA}

\author[0000-0002-8585-0084]{Colleen A. Wilson-Hodge}
\affiliation{NASA Marshall Space Flight Center, Huntsville, Al 35812, USA}

\begin{abstract}

GW230529 is the first compact binary coalescence detected by the LIGO--Virgo--KAGRA collaboration with at least one component mass confidently in the lower mass-gap, corresponding to the range 3--5$M_{\odot}$. If interpreted as a neutron star--black hole merger, this event has the most symmetric mass ratio detected so far and therefore has a relatively high probability of producing electromagnetic (EM) emission. However, no EM counterpart has been reported. At the merger time $t_0$, \textit{Swift}--BAT and \textit{Fermi}--GBM together covered 100$\%$ of the sky. Performing a targeted search in a time window $[t_0-20 \text{s},t_0+20 \text{s}]$, we report no detection by the \textit{Swift}--BAT and the \textit{Fermi}--GBM instruments. Combining the position-dependent $\gamma-$ray flux upper limits and the gravitational-wave posterior distribution of luminosity distance, sky localization and inclination angle of the binary, we derive constraints on the characteristic luminosity and structure of the jet possibly launched during the merger. Assuming a \textit{top--hat} jet structure, we exclude at 90$\%$ credibility the presence of a jet which has at the same time an on-axis isotropic luminosity $\gtrsim 10^{48}$ erg s$^{-1}$, in the bolometric band 1 keV--10 MeV, and a jet opening angle $\gtrsim 15$ deg. Similar constraints are derived testing other assumptions about the jet structure profile. Excluding GRB 170817A, the luminosity upper limits derived here are below the luminosity of any GRB observed so far.
\end{abstract}


\section{Introduction} \label{sec:intro}
Since operations began in 2015, the LIGO--Virgo--KAGRA (LVK) detector network \citep{2015CQGra..32g4001L,2015CQGra..32b4001A,2021PTEP.2021eA101A} has successfully identified gravitational-waves (GWs) consistent with binary black hole (BBH) mergers, binary neutron star (BNS) mergers, and neutron star--black hole (NSBH) mergers \citep{GWTC2Abbott2021PhRvX11b1053A,Abbott2023PhRvXGwtc3}. NSBH mergers, as BNS mergers, can be accompanied by a short $\gamma-$ray burst (GRB) and/or a kilonova emission, with the kilonova due to the radioactive decay of heavy elements in the neutron-rich ejecta launched by the merger \citep{Rosswog2005ApJ...634.1202R,Tanaka2014ApJ...780...31T,2019LRR....23....1M,Gompertz2020ApJ, 2018IJMPD..2742004C,2022ApJ...936L..10Z,GompertzMMM2023MNRAS.526.4585G}. The intrinsic brightness of the EM emission strongly depends on the amount of mass released before the neutron star enters the innermost stable circular orbit of the black hole, impacted in turn by the mass ratio and the magnitude and orientation of the spins of the two objects \citep{2015PhRvD..92b4014K,2018PhRvD..98h1501F,2020PhRvD.101j3002K}. Specifically, low-mass BHs, more compact NSs and high prograde spins are all factors that favor tidal disruption and make the NSBH candidate potentially EM bright. Moreover, additional conditions are required for a NSBH merger to successfully launch a relativistic jet \citep{2020EPJA...56....8B,2023arXiv231016894C}. On average, NSBH mergers have a smaller probability of having enough material to be collimated in a jet, compared to BNS mergers, though they have the advantage that the polar axis is not expected to be baryon polluted, facilitating the propagation and acceleration of the relativistic outflow \citep{2015PhRvD..92d4028K}.

Among all the compact binary merger candidates found by the LVK collaboration up to the time of writing, GW191219$\_$163120 and GW200115$\_$042309 are confidently classified as NSBH mergers with $p_{\rm astro}>$0.5 \citep{Abbott2021ApJ915NSBH,Abbott2023PhRvXGwtc3,O3CatUpperLimits}\footnote{At the time GW200105$\_$162426 was also identified as a NSBH candidate. Subsequent analysis found this trigger to be of marginal interest with $p_{\rm astro}<$0.5 \citep{Abbott2023PhRvXGwtc3}.}. Focusing on GW200115$\_$042309, this merger was found to have a 27--30\% probability that the primary component falls within the range of lower mass gap depending on the choice of spin prior \citep{Abbott2021ApJ915NSBH}. Further analysis indicated that $<10^{-6} M_{\odot}$ was ejected from this merger at $>$99\% credibility, based on calculations of dynamical ejecta and mass loss due to disruption \citep{Abbott2021ApJ915NSBH}.  This is consistent with the non-detection of electromagnetic emission accompanying these candidates \citep{Simone2021ApJ...923L..32D,Zhu2021ApJ...921..156Z, Anand2021NatAs...5...46A,O3CatUpperLimits}. 




In addition, GW190814 and GW200210$\_$092254 are particularly noteworthy, since the inferred secondary mass of 2.59$^{+0.08}_{-0.09}$ M$_{\odot}$ (for GW190814) and  2.6$^{+0.1}_{-0.1}$ M$_{\odot}$ (for GW200210$\_$092254) indicates that the source could have been an extremely massive NS or a BH \citep{190814-2020ApJ...896L..44A,GWTC22021PhRvX..11b1053A,Abbott2023PhRvXGwtc3}. Neither of these NSBH candidates were found to have associated EM emission in their follow-up observations \citep{DObie2019ApJ...887L..13D,Watson2020MNRAS.492.5916W,Thakur2020MNRAS.499.3868T,Alexander2021ApJ...923...66A,Kilpatrick2021ApJ...923..258K,Fletcher_2024}

Given the lack of a direct association, there is still no consensus about the fraction of short GRBs potentially driven by NSBH mergers. However, \cite{2023RNAAS...7..136B} find an upper limit of 20 Gpc$^{-3}$ yr$^{-1}$ on the rate of GRBs with NSBH progenitors based on the population of mergers observed in gravitational waves. Whether there exists a sub-population of short GRBs whose distinctive properties could distinguish these merger channels remains unsettled. Several studies have tried to explore such a dichotomy deeper by examining samples of short GRBs through some of their observable properties such as duration, energetics, kilonovae emission, locations, and offsets from host galaxies \citep{Troja2008MNRAS.385L..10T,Gompertz2020ApJ,2023ApJ...949L..22D}. These studies suggest that short GRBs with extended emission could indeed be classified as a unique population that potentially arises from NSBH mergers. Moreover, depending on the amount of disk mass and the accretion timescale, NSBH mergers can power GRBs whose duration extends well beyond the canonical 2 seconds, as observed for the two merger-driven long duration bursts: GRB 211211A \citep{2022Natur.612..223R,2022Natur.612..228T,2022Natur.612..236M,2022Natur.612..232Y} and GRB 230307A \citep{2024Natur.626..737L,2024Natur.626..742Y}.

The paucity of significant detections of NSBH mergers in previous LVK observing runs limits our understanding of the NSBH population, the associated merger rate and probability of producing a bright EM emission. Detecting these events through their $\gamma-$ray emission, or setting sensitive upper limits will allow us to derive constraints on the parameters associated with the various predicted GRB jet emission models and subsequently allow us to potentially differentiate better the BNS and NSBH merger channels \citep{2020LRR....23....4B}. Moreover, since the quality of GW data is not always good enough to confirm or exclude with confidence the presence of a NS in a compact binary merger, the detection of an EM counterpart would be the smoking gun to prove the presence of baryonic matter surrounding the final remnant object. Past multi-messenger studies have already demonstrated how the combination of GW data and $\gamma-$ray observations could shed light on the association between BNS/NSBH mergers and short GRBs, as well as on the jet properties, including the opening angle and structure profile \citep{2020ApJ...891..124H,2020ApJ...895..108F,2020ApJ...893...38B,2023ApJ...954...92H}.

In this work we examine the non-detection of $\gamma-$ray emission from the gravitational wave trigger GW230529$\_$181500, hereafter GW230529 \citep{2024arXiv240404248T}, based on the monitoring from \textit{Swift}--BAT and \textit{Fermi}--GBM. We summarize the GW properties of this NSBH event in Section \ref{sec:gw}. In Section \ref{sec:bat} and \ref{sec:gbm} we discuss the \textit{Swift}--BAT and \textit{Fermi}--GBM observations and the targeted search analysis that was conducted around the trigger time of the event. Using joint BAT--GBM flux upper limits (Section \ref{sec:joint_bat_gbm}), we further discuss methods adopted to place constraints on the jet luminosity and opening angle (Section \ref{sec:jet-constraints}). We discuss the simulations in Section \ref{sec:sim} and present our results in Section \ref{sec:results}.

\section{Information from the GW analysis} \label{sec:gw}

\cite{2024arXiv240404248T} report that GW230529 is a compact binary coalescence whose primary is a neutron star of mass $1.4_{-0.2}^{+0.6}$ $M_{\odot}$, merging with a compact object of mass $3.6_{-1.2}^{+0.8}$ $M_{\odot}$, at 90$\%$ credibility. In the hypothesis of a highly spinning secondary component, and marginalizing over the component mass distributions and the NS equation of state (EoS), there is $<10\%$ probability that the NS is tidally disrupted. The corresponding ejected mass due to the tidal disruption of the NS is $<0.052M_{\odot}$ at 99$\%$ credibility. This limit is quite informative, since, under optimal conditions, NSBH mergers can eject up to $\sim$0.1--0.3 $M_{\odot}$ (e.g., \citealt{2015PhRvD..92d4028K}). With the detection of GW230529 we now know that events with low mass ratios exist, increasing the probability that the NSBH class of events can be EM bright. Indeed, with the inclusion of GW230529 in the sample of NSBH detected so far, \cite{2024arXiv240404248T} derive that up to $18\%$ of the population can power an EM counterpart, a factor of three larger compared to the same analysis that excluded GW230529. The authors also report that, due to the higher probability of having a tidal disruption of the NS, the upper bound on the rate of GRBs coming from NSBH progenitors increases to a value of 23 Gpc$^{-3}$ yr $^{-1}$, based on the methodology developed by \cite{2023RNAAS...7..136B}.

The uncertainty in the sky localization is remarkably large, covering almost the totality of the sky ($\simeq 24100$ deg$^2$ at 90$\%$ credibility). The luminosity distance is constrained in the range $D_L=201_{-96}^{+102}$ Mpc ($z\sim0.045$). The posterior distribution of the inclination angle of the binary is uninformative, being very close to the detection prior, which follows the Malmquist bias (see, e.g., \citealt{2011CQGra..28l5023S}). The posterior samples required for this analysis are downloaded from the LVK data release repository \citep{ligo_scientific_collaboration_2024_10845779}. Here we consider the posterior samples labeled as \texttt{Combined}$\_$\texttt{PHM}$\_$\texttt{highSpin}, which are the combination of posteriors obtained independently using the \texttt{IMRPhenomXPHM} \citep{Pratten2021PhRvD} and \texttt{SEOBNRv4PHM} \citep{Ossokine2020PhRvD} waveforms, in the assumption of high-spin prior for both components. For consistency, throughout the paper we use the corresponding sky localization map \texttt{skymap}$\_$\texttt{combined}$\_$\texttt{PHM}$\_$\texttt{high}$\_$\texttt{spin.fits}.

An independent follow-up analysis performed by \cite{2024arXiv240410596Z} finds that, considering APR4 \citep{Akmal1998PhysRevC} and DD2 \citep{PhysRevC.81.015803} as EoS, the probability that in the merger the NS is tidally disrupted is 12.8$\%$ and 63.2$\%$, respectively. Moreover, combining this event with population models of NSBH mergers, the authors find that $>90\%$ of tidally disrupted NSBHs come from the sub-class of events where the primary mass is in the $[3-5]M_{\odot}$ mass range. From the detectability perspective, due to the small mass ratio between BH and NS, the higher chance of having matter around the central remnant implies that the rate of detectable GRBs (within 300 Mpc) formed by NS+[3-5] $M_{\odot}$ systems is at least a factor 10 larger than the rate of GRBs formed by NS+[$>5$]~$M_{\odot}$ systems. The factor is even larger for the detectability of the kilonova component. 

Numerical simulation performed by \cite{2024arXiv240506819M} in the mass regime of GW230529 reveal the possible emergence of a low-mass accretion disk, capable of powering a short GRB \citep{2023ApJ...958L..33G}. Therefore, it is evident that events like GW230529 are, among NSBH mergers, the most promising candidates to have an associated EM emission, in the form of a GRB and/or kilonova emission. Any constraint on the EM nature of these systems is notably valuable to obtain deeper insights about the connection between the merger parameters (component masses, spins and inclination angle) and the capability to power GRB/kilonova emission.

\begin{figure*}
    \centering
    \includegraphics[width=0.48\textwidth]{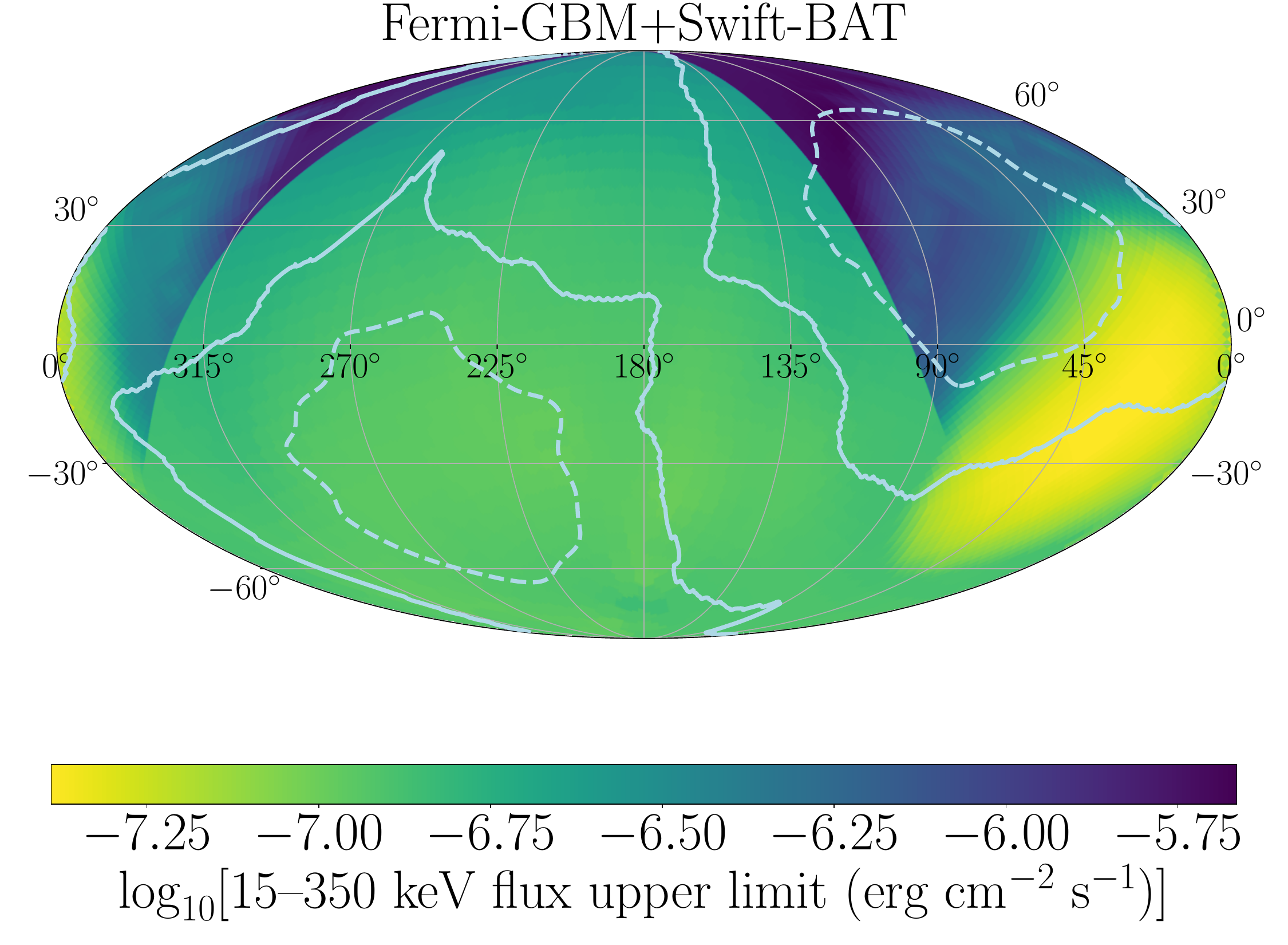} \quad
    \includegraphics[width=0.48\textwidth]{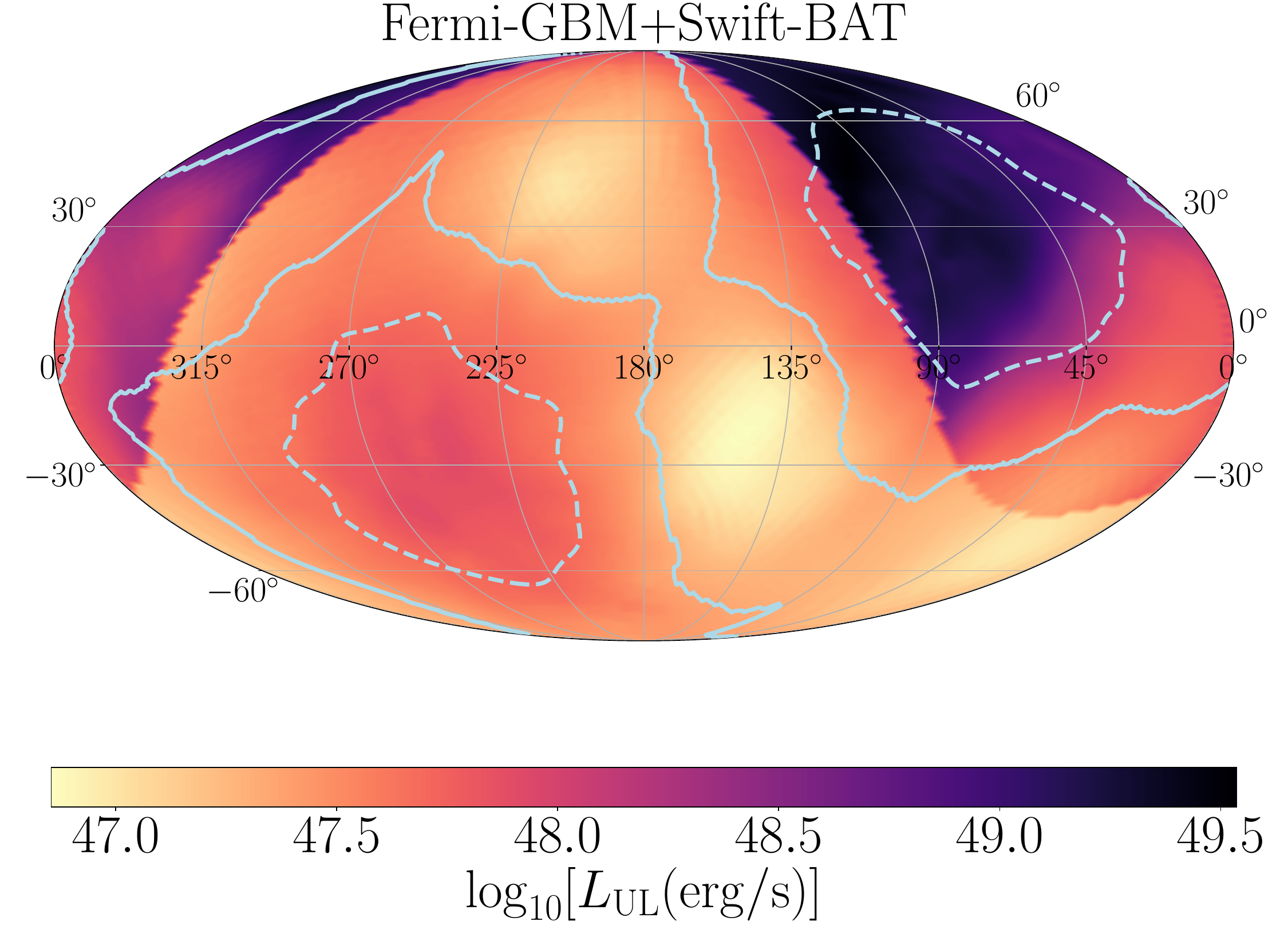}\\

    \includegraphics[width=0.7\textwidth]{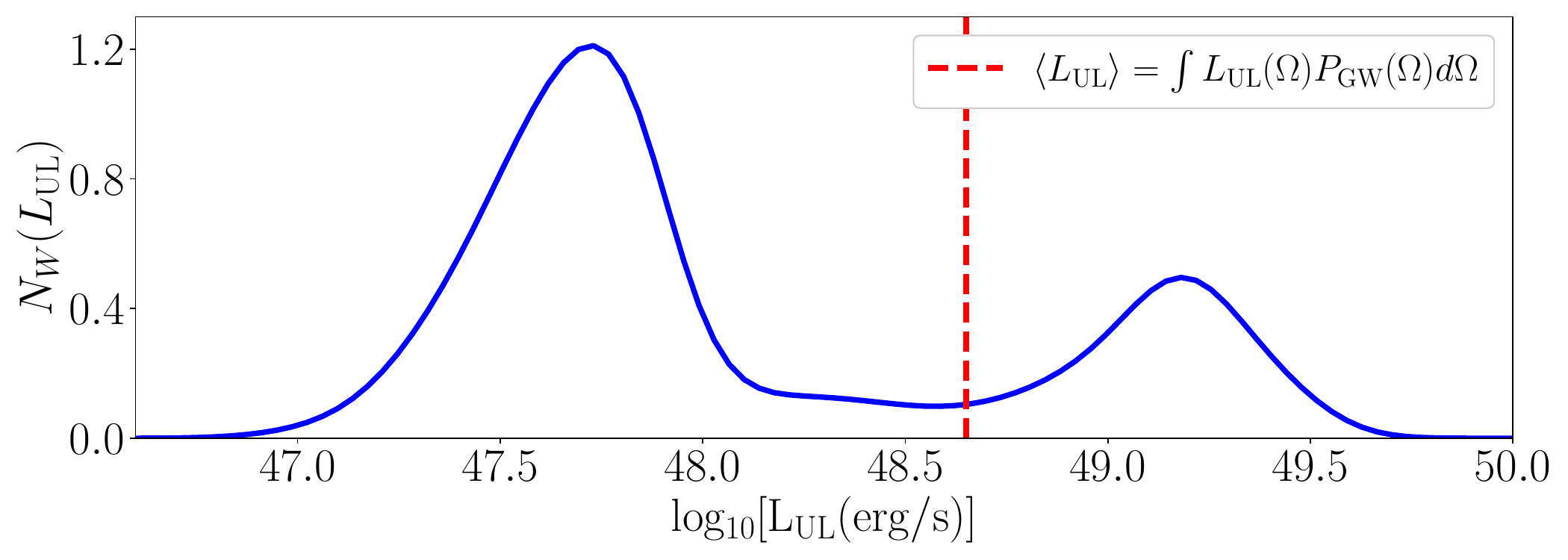}
    \caption{Upper left panel: joint \textit{Swift}--BAT+\textit{Fermi}--GBM flux upper limit sky map in the 15--350 keV band. The upper limits are relative to a time scale of 1 second, computed at 5$\sigma$ confidence level. For the individual contributions from the two telescopes, see Fig~\ref{fig:ul}. Upper right panel: joint \textit{Swift}--BAT+\textit{Fermi}--GBM sky map of the bolometric luminosity upper limit. The luminosity is computed in the rest frame energy range 1 keV--10 MeV. In both upper panels, the solid and dashed lines are the GW localization contours at 90$\%$ and 50$\%$ credibility, respectively. Lower panel: weighted distribution of the luminosity upper limit, where the weight is given by the GW probability density of the single pixel. The vertical dashed line represents the weighted average over the sky of the luminosity upper limit.}
    \label{fig:ul_lum}
\end{figure*}

\section{\textit{S\MakeLowercase{wift}}--BAT observation} \label{sec:bat}

The Neil Gehrels \textit{Swift} Observatory \citep{Gehrels2004}, launched in 2004 has successfully detected and localized $>$1600 GRBs in its almost 20 years of operations. This is thanks in most part to its GRB monitor, \textit{Swift}--BAT \citep{Barthelmy2005}. \textit{Swift}--BAT is a wide field of view ($\sim$2 sr), coded mask imager that is sensitive to $\gamma-$rays in the 15--350 keV energy range. By correlating the spatial distribution of detected counts across \textit{Swift}--BAT's detector plane with the pattern of its coded mask, an image of the sky can be created, and sources can be localized within a few arcminutes. 

At the trigger time of GW230529, \textit{Swift} was in a pointed observation and in a good data taking mode. There were no onboard GRB triggers within an hour of GW230529. 

During the full period of O4 the Gamma--ray Urgent Archiver for Novel Opportunities (GUANO, \citealt{GUANOTohuvavohu2020}) enables the downlink of \textit{Swift}--BAT data for all the GW triggers with a false alarm rate $<2$ day$^{-1}$. For GW230529 GUANO downloaded 200 s of event data around the trigger time. We performed a targeted search with the NITRATES pipeline \citep{NITRATES2022} in a temporal window $[t_0-20\,s, t_0+20\,s]$, where $t_0=$2023--05--29T18:15:00 UTC is the GW trigger time. The analysis searched for impulsive $\gamma-$ray emission over eight different time scales: 0.128~s, 0.256~s, 0.512~s, 1.024~s, 2.048~s, 4.096~s, 8.192~s, and 16.384~s. The most significant candidate identified by the search has a $\sqrt{\rm TS}=5.7$, which corresponds to a false alarm rate $>$ 0.01 s$^{-1}$. $\sqrt{\rm TS}$ is defined in \cite{NITRATES2022} and gives the statistical significance of the detection. With these search results we find no significant evidence of a possible $\gamma-$ray signal associated with GW230529. With a non-detection, upper limits are set with the method introduced in \cite{O3CatUpperLimits}, but with added responses for positions outside the coded field of view to make the upper limits cover the full Earth unocculted sky.

We derive flux upper limits considering a temporal bin of 1 s with a confidence level of 5$\sigma$, unless otherwise stated. The choice of the temporal bin is made according to the theoretical expectation of the typical duration of GRBs originated by NSBH systems. We consider three spectral templates, consisting of a Band function \citep{band1993ApJ...413..281B}, with a peak energy of 75~keV (soft template), 230 keV (normal template) and 1500~keV (hard template), while the low- and high-energy photon indices are fixed to $\alpha=-1$ and $\beta=-2.3$, respectively. As discussed in Section \ref{sec:jet-constraints}, we have no strong priors on the inclination angle of the binary; therefore, it would be unmotivated to assume a peak energy typical of on-axis short-hard GRBs. Therefore, for each pixel of the sky map, we consider the spectral template that produces the largest value of the flux upper limit. In this way, the adopted upper limit is the most conservative one, without impacting the solidity of our conclusions.

If $P_{GW}$ is the GW sky localization probability, $\epsilon_{\rm cov}$ the probability that the GW event occurred in a position of the sky not occulted by the Earth and $\epsilon_{\rm in \, FoV}$ the probability that the GW event occurred inside the BAT coded field of view, we have 
\begin{equation}
\label{ea}
    \epsilon_{\rm cov}=1-\int_{\Omega \in \Omega_{\Earth}} P_{GW}(\Omega) d\Omega,
\end{equation}
and
\begin{equation}
\label{infov}
    \epsilon_{\rm in \, FoV}=\int_{\Omega \in \Omega_{\rm BAT}} P_{GW}(\Omega) d\Omega,
\end{equation}
where $\Omega$ denotes the sky coordinates. We obtain the values 
$\epsilon_{\rm cov}=63\%$ and $\epsilon_{\rm in \, FoV} = 15\%$. 

\section{\textit{F\MakeLowercase{ermi}}--GBM observation} \label{sec:gbm}
\textit{Fermi}--GBM consists of 12 sodium iodide (NaI) and 2 bismuth
germanate (BGO) detectors that cover
the full sky, unocculted by the Earth  \citep{Meegan2009}. 
The NaI detectors are sensitive to photons in the energy range 8~keV to 1000~keV, and the BGO detectors observe the 200~keV to 40~MeV energy range. The flight software onboard \textit{Fermi}--GBM triggers on any event with a flux of $\gamma-$rays at a level greater than a threshold (typically  $\sim 4.5$--$5\sigma$) above the background rate in at least two NaI detectors \citep{von_Kienlin_2020}. In more than 15 years of operations, \textit{Fermi}--GBM has detected more than 3750 GRBs with onboard triggers. 

In addition to trigger data, \textit{Fermi}--GBM gives us high time resolution (2 $\mu$s) data in the form of continuous time tagged events (CTTEs) over the energy range from 8~keV to 40~MeV. The CTTE data allow us to search for GRBs below \textit{Fermi}--GBM’s on-board trigger threshold using ground-based computing resources \citep{Blackburn+15targeted, Goldstein+19targeted}. The most sensitive search method, referred to as the \textit{Targeted Search} pipeline, is a likelihood-based approach for multimessenger follow-up observations \citep{Blackburn+15targeted}. More details about the \textit{Targeted Search} pipeline can be found in \citet{Blackburn+15targeted, Goldstein+19targeted, Hamburg_2020, Fletcher_2024}.

At the merger time of GW230529, a fraction $\epsilon_{\rm cov}=62\%$ of the GW localization was visible to \textit{Fermi}--GBM. There was no on-board trigger near this time. We, therefore, performed a search for GRB-like emission with \textit{Targeted Search} from $\pm$20 s around the merger time. The scan was repeated for nine characteristic emission timescales that increased by factors of 2 from 64 ms to 16.384 s using the three characteristic GRB spectra from Table 3 of \cite{Fletcher_2024}. No significant counterparts were identified by this search.

 As we did not find a significant counterpart in \textit{Fermi}--GBM for GW230529, we compute the $\gamma-$ray flux upper limits as a function of sky position using the pipeline \textit{Targeted Search}. We do this by computing the upper limits, with a 5$\sigma$ confidence level, separately for each spectral template. Similar to \textit{Swift}--BAT, we then choose the least constraining limit from all three spectral templates for each position to obtain a final upper limit, which is less sensitive to the individual spectral shapes.  

\section{Joint BAT+GBM flux and luminosity upper limits} \label{sec:joint_bat_gbm}

We combine the flux upper limits from \textit{Swift}--BAT and \textit{Fermi}--GBM, indicated as $\rm UL_{BAT}(\Omega)$ and $\rm UL_{GBM}(\Omega)$, respectively, producing a joint flux upper limit map defined as
\begin{equation}
\label{joint_ul}
    \rm UL_{joint}(\Omega) = min[UL_{BAT}(\Omega), \rm UL_{GBM}(\Omega)].
\end{equation}
The map is shown in the upper left panel of Fig.~\ref{fig:ul_lum}, superimposed on the contour levels of the GW sky localization. The flux upper limits are reported in the 15--350 keV energy band. Lighter colored regions indicate more sensitive upper limits. The solid and dashed lines are the GW sky localization contours at $90\%$ and $50\%$ credibility, respectively. The single \textit{Swift}--BAT and \textit{Fermi}--GBM flux upper limits maps are shown in the Appendix (Fig.~\ref{fig:ul}), where the white region indicates the fraction of the sky covered by the Earth. 

We also compute a sky-dependent bolometric luminosity upper limit in the rest-frame energy band 1~keV--10~MeV, as
\begin{equation}
    L_{\rm UL} (\Omega) = 4 \pi \bar{D}_L^2(\Omega) k(\Omega) \rm UL_{joint}(\Omega),
\end{equation}
where:
\begin{itemize}
    \item $\bar{D}_L(\Omega)$ is the mean luminosity distance as a function of the sky position. This in turn is derived as
    \begin{equation}
        \bar{D}_L (\Omega) = \int D_L P_{D_L} (D_L|\Omega)dD_L,
    \end{equation}
    with $P_{D_L} (D_L|\Omega)$ the sky position-dependent posterior distribution of the luminosity distance.
    \item $k(\Omega)$ is the k-correction defined as
    \begin{equation}
  k(\Omega)=  \frac{I[1 \,\mathrm{ keV}/(1+z(\Omega)),10 \, \mathrm{ MeV}/(1+z(\Omega))]}{I[15 \,\mathrm{ keV}, 350 \,\mathrm{ keV}]},
\end{equation}
where 
\begin{equation}
  I[a,b]=\int_{a}^{b} E \frac{\mathrm{d} N}{\mathrm{~d} E}(\Omega) \mathrm{d} E,  
  \end{equation}
and d$N$/d$E(\Omega)$ is the assumed photon spectrum. For the computation of $z(\Omega)$ we adopt the values of \cite{2020A&A...641A...6P}, namely $h=0.68, \Omega_{\Lambda}=0.69, \Omega_{m}=0.31$. 
\end{itemize}

The sky map of $L_{\rm UL} (\Omega) $ is shown in the upper right panel of Fig.~\ref{fig:ul_lum}. In the lower panel of Fig.~\ref{fig:ul_lum} we show the distribution of $L_{\rm UL} (\Omega) $ over all the sky pixels. The distribution $N_{W}(L_{\rm UL})$ is obtained considering the collection $L_{\rm UL} (\Omega_i) $ for each pixel $i$ and weighted by the GW sky probability density $P_{GW}(\Omega_i)$. $N_{W}(L_{\rm UL})$ shows a double peak, the lower one being due to the more constraining upper limits obtained by \textit{Fermi}--GBM and \textit{Swift}--BAT (inside the field of view), while the higher one due to the less constraining values obtained outside the  \textit{Swift}--BAT field of view. The vertical dashed line in 
the lower panel of Fig.~\ref{fig:ul_lum} reports the weighted average of $L_{\rm UL} (\Omega) $, defined as
\begin{gather}
 \langle L_{\rm UL} \rangle =\int L_{\rm UL} (\Omega) P_{GW} (\Omega) d\Omega \\ \notag 
=2.5 \times 10^{48} \rm erg/s.
\end{gather}

\section{Constraining the $\gamma-$ray emission from GW230529} \label{sec:jet-constraints}

\begin{figure}
    \centering
    \includegraphics[width=1.0\columnwidth]{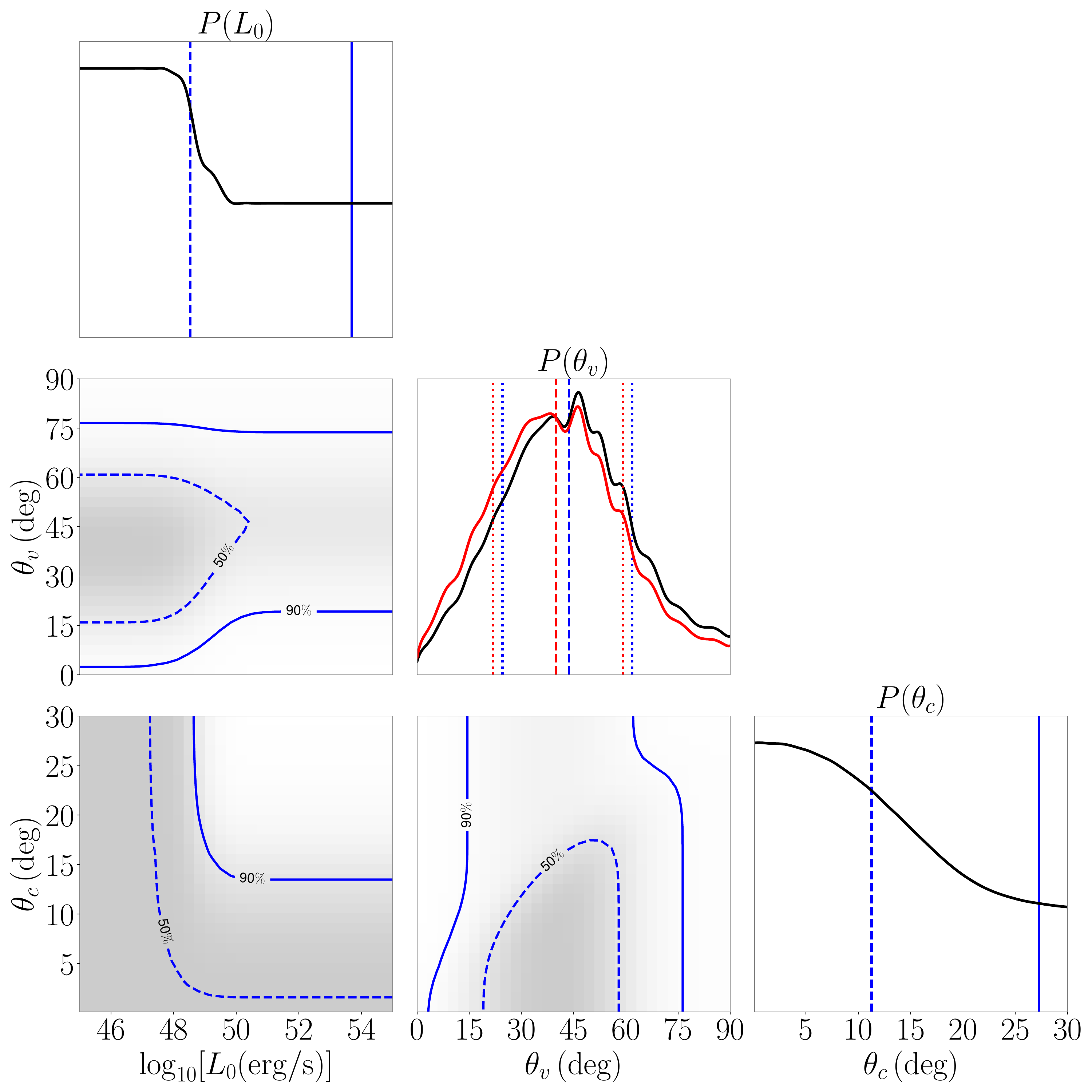}
    \caption{Corner plot of the posterior distribution for the parameters $L_0$, $\theta_c$ and $\theta_v$, assuming a \textit{top--hat} jet structure. In the off-diagonal plots, the dashed and solid lines indicate the $50\%$ and $90\%$ credible regions. In the middle panel $P(\theta_v)$, we report in red the prior $\pi(\theta_v)$ derived from the GW analysis. The vertical dotted lines identify the 1$sigma$ credible interval ($16^{\rm th}$ and $84^{\rm th}$ percentile), while the dashed lines the median ($50^{\rm th}$ percentile). For the other posteriors, $P(L_0)$ and $P(\theta_c)$, the vertical dashed and solid lines report the $50^{\rm th}$ and $90^{\rm th}$ percentiles of each posterior, respectively. The gray scale in the off-diagonal plots indicates the posterior density, where darker corresponds to larger values. The excluded regions are outside the gray areas.}
    \label{corner_tophat}
\end{figure}

\begin{figure}
    \centering
    \includegraphics[width=1\columnwidth]{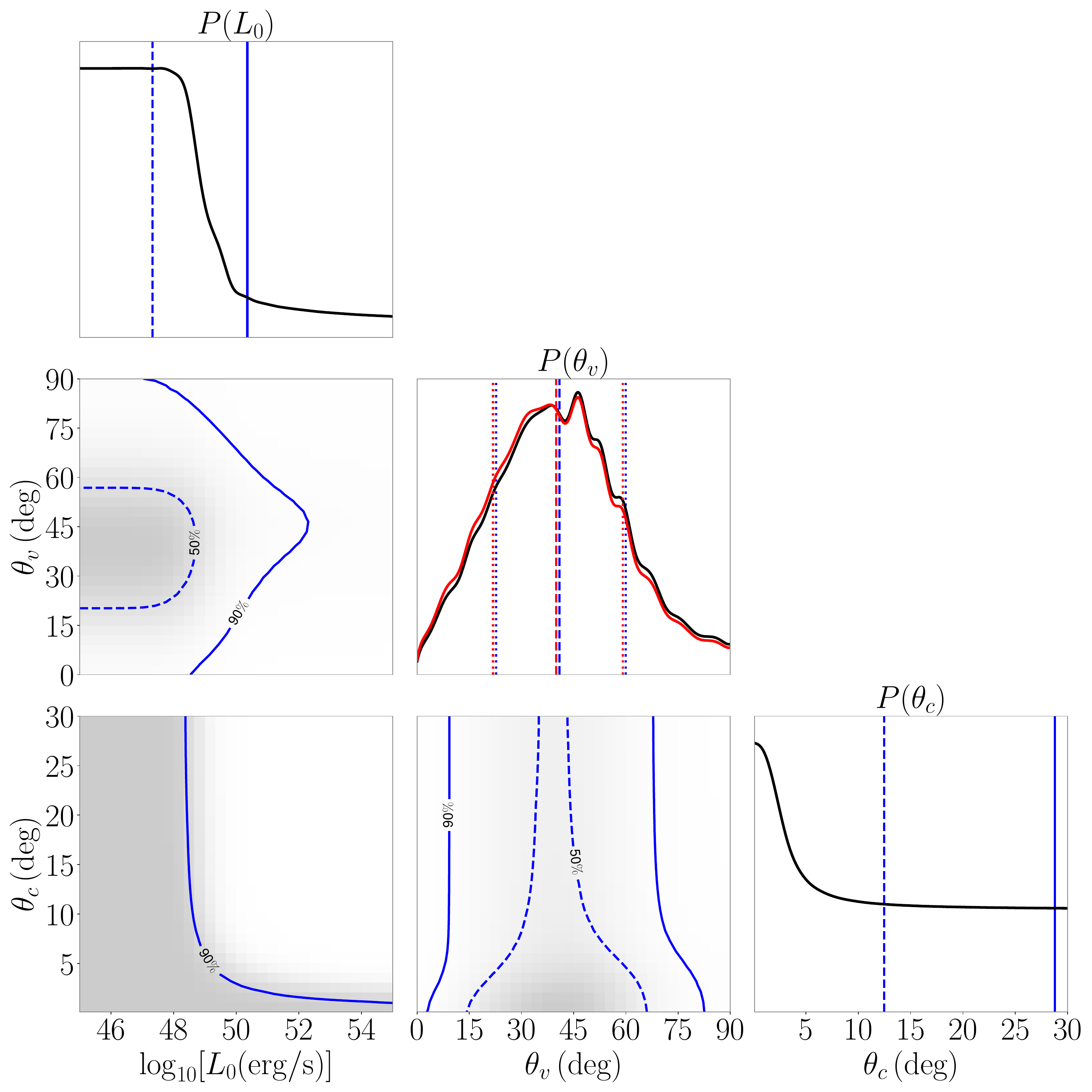} 
    \caption{Same as Fig.\ref{corner_tophat}, but assuming a \textit{Gaussian} profile for the jet structure.}
    \label{corner_gauss}
\end{figure}

The non-detection of any $\gamma-$ray signal coincident with GW230529 can be interpreted as follows. In the NSBH merger scenario, one possibility is that the NS is entirely swallowed by the BH and the absence of any accretion material prevents the formation and launch of a relativistic jet, necessary for the GRB emission. On the other hand, even if a jet is launched, it might be too off-axis to produce any detectable $\gamma-$ray signal. A third option is that the amount of material accreted onto the central engine is so scarce that the resulting jet, even if aligned with the observer, is not powerful enough to produce a detectable $\gamma-$ray signal. Additionally, since there are no constraints on the tidal deformability of the secondary component, it is not excluded that GW230529 is an exotic merger of unusually light BHs, possibly of primordial origin (e.g., \citealt{2022arXiv221105767E,2024arXiv240405691H}). In this last case, unless the BBH merger is embedded in a very dense environment \citep{2023ApJ...942...99G}, no EM counterpart is expected. 

The aim of this section is to exploit the non-detection upper limits of \textit{Swift}--BAT and \textit{Fermi}--GBM to infer possible constraints on the typical luminosity of the jet. As expected from theoretical simulations and confirmed by observations, the jets of GRBs have an angular structure that can deviate from the top-hat approximation \citep{2002MNRAS.332..945R,2003ApJ...591.1086G,2017MNRAS.472.4953L,2017ApJ...848L..12A,2017Natur.551...71T,2018MNRAS.478L..18T,2019MNRAS.489.1919T,2019Sci...363..968G,2020ApJ...896..166R,2022Galax..10...93S}. In general, the jet structure can be described by a core with opening angle $\theta_c$, where the energy radiated per unit solid angle $E(\theta_v)$ is approximately constant, and an off-core region, where the radiated energy drops off rapidly as a function of the inclination angle. Here we construct a simple toy model to describe the analytical form of the apparent structure of the jet luminosity, namely as measured by an observer located at an inclination angle $\theta_v$ from the jet axis. Formally, the apparent structure of luminosity $L(\theta_v)$ does not necessarily follow the same profile of $E(\theta_v)$ (see Appendix \ref{lum_app} for further details).
We parametrize the apparent structure of the jet luminosity as:
\begin{equation}
    L(\theta_v) = L_0 l(\theta_v),
\end{equation}
where $L_0=L(\theta=0)$, while
\begin{equation}
l(\theta_v)= \begin{cases}
\sim 1, & \theta < \theta_c \\
f(\theta_v), & \theta > \theta_c
\end{cases}.
\end{equation}

In the following, we focus on three possible profiles of the jet structure:
\begin{enumerate}
    \item \textit{Top--hat}:
    \begin{equation}
        l(\theta_v)= \begin{cases}
 1, & \theta < \theta_c \\
0, & \theta > \theta_c
\end{cases}.
    \end{equation}
    \item \textit{Gaussian}:
        \begin{equation}
        l(\theta_v) = \exp \left[-\frac{\theta_v^2}{2\theta_c^2}\right].
    \end{equation}
        \item \textit{Power law}:
        \begin{equation}
        l(\theta_v) = \dfrac{1}{1+\left(\frac{\theta_v}{\theta_c}\right)^s}.
    \end{equation}

    \item \textit{Two components}: 
    \begin{equation}
        l(\theta_v)= \begin{cases}
 1, & \theta < \theta_c \\
L_{\rm off}/L_0, & \theta > \theta_c
\end{cases}.
\end{equation}
    This structure consists of a \textit{top--hat} plus an isotropic component outside the jet core.

\item \textit{Isotropic}:
\begin{equation}
        l(\theta_v)=1.
\end{equation}
For completeness, we explore also the possibility that this GW merger did not produce a jetted GRB-like emission, but a more isotropic component, whose nature could be associated with the precursor emission and/or the extended emission, typically observed in merger-driven GRBs. Since this kind of emission can occur on longer timescales ($\sim$ tens of seconds), we consider an upper limit computed on the timescale of 16.384~s, the longest available from the templates.
\end{enumerate}

The apparent luminosity $L(\theta_v)$ has to be interpreted as the rest-frame luminosity computed in the bolometric energy range 1 keV--10 MeV. Moreover, the luminosity defined here does not need the definition of a spectral model. As explained more in detail in Appendix \ref{lum_app}, given a limit on $L(\theta)$, one can derive a constraint on the spectrum normalization once the spectral shape is known. This is in line with the derivation of a flux upper limit using a set of different spectral templates.

\section{Simulation setup} \label{sec:sim}
\begin{figure}
    \centering
    \includegraphics[width=1\columnwidth]{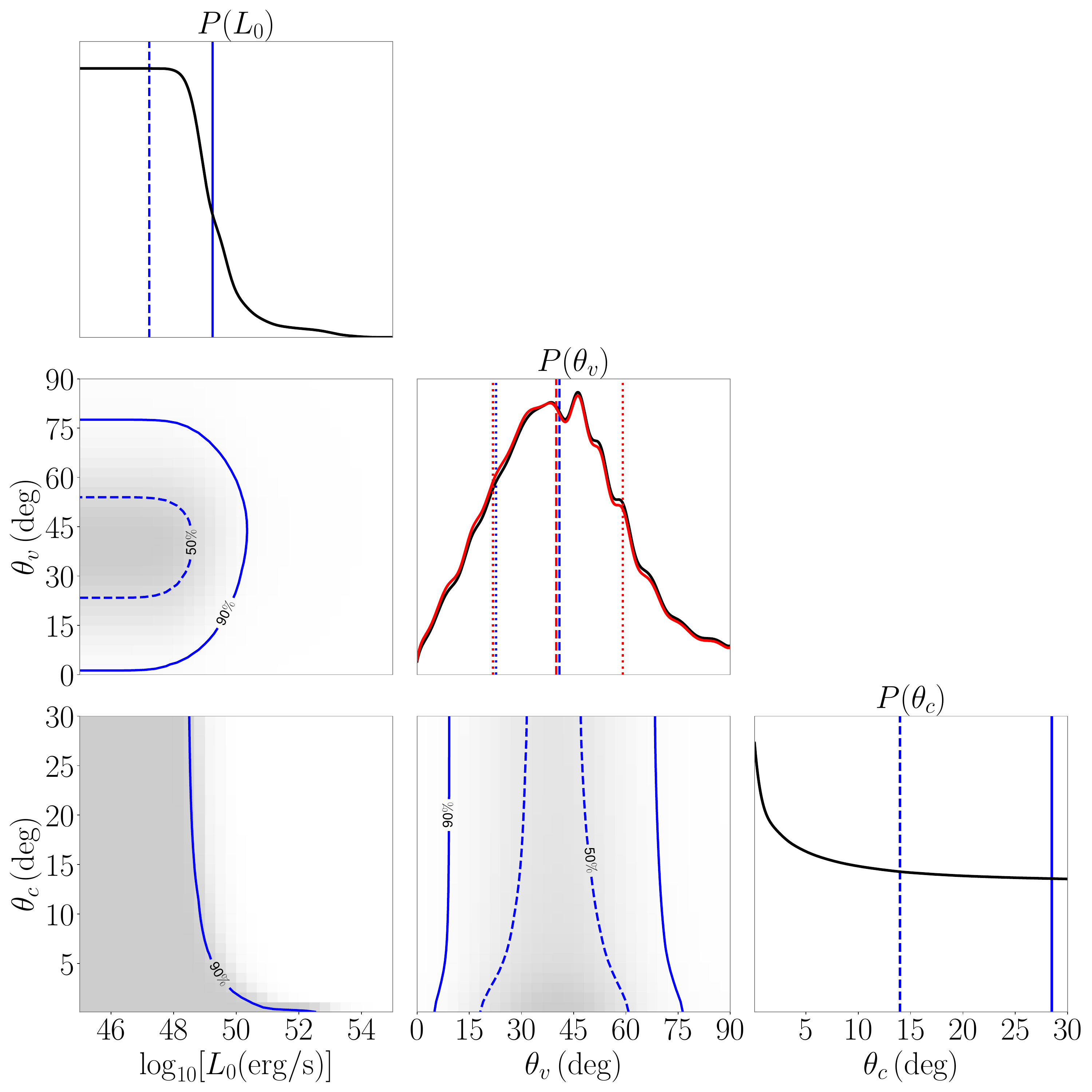}\\
    \includegraphics[width=1\columnwidth]{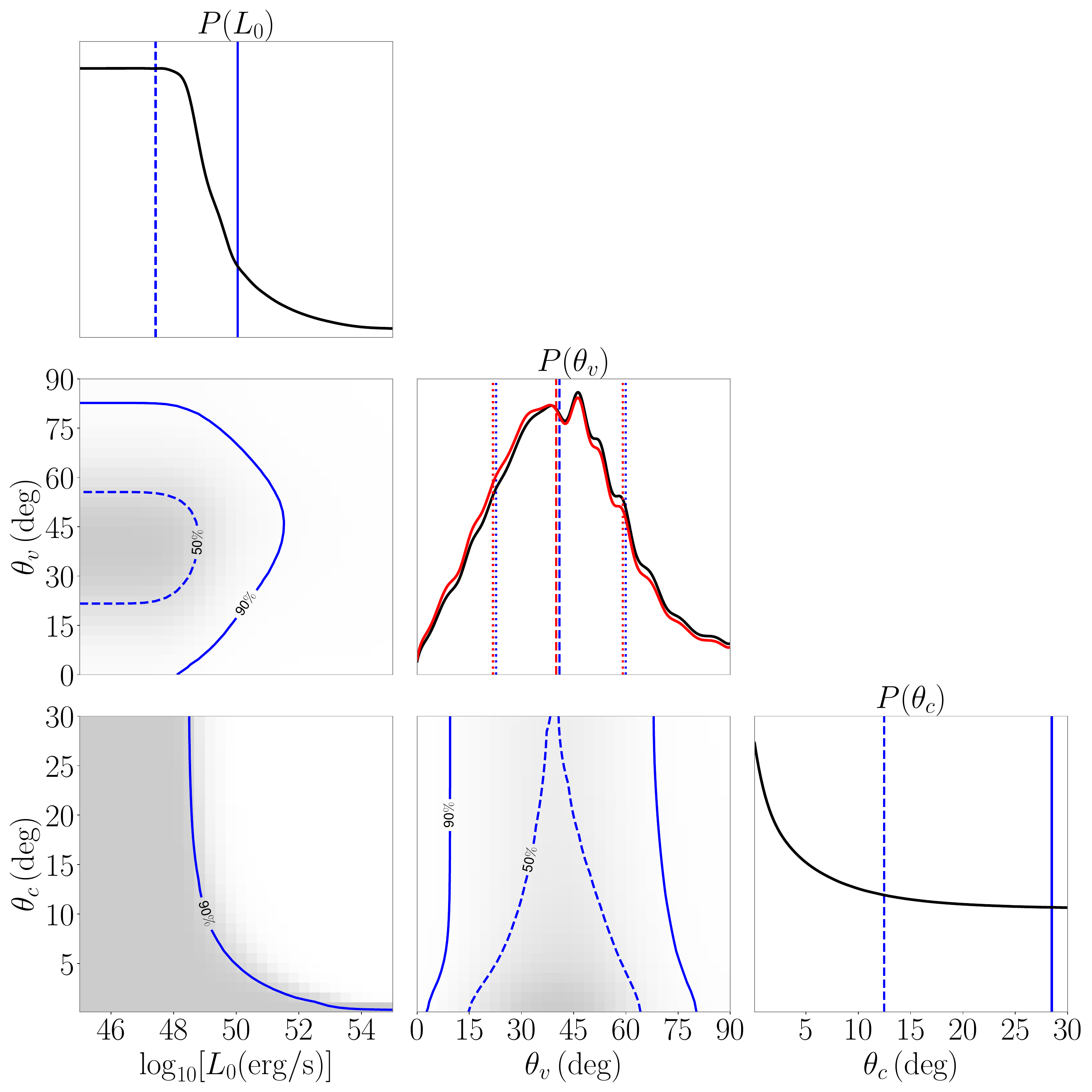}
    \caption{Same as Fig.\ref{corner_tophat}, but assuming a power law profile for the jet structure, with an off-axis slope $s=2$ (top) and $s=4$ (bottom).}
    \label{corner_pl}
\end{figure}

The aim of this section is to set up a simulation which combines the posterior distribution of the GW parameters with the sky position dependent upper limits of \textit{Swift}--BAT $\rm UL_{BAT}(\Omega)$ and \textit{Fermi}--GBM $\rm UL_{GBM}(\Omega)$, in order to infer constraints about the parameters of the jet structure. As specified in Section~\ref{sec:joint_bat_gbm}, for each pixel of the sky we consider an upper limit 
\begin{equation}
    \rm UL(\Omega)=\rm UL_{joint}(\Omega),
\end{equation}
defined by Eq.~\ref{joint_ul}. To derive constraints on the luminosity in the bolometric range 1 keV--10 MeV, we apply for each pixel of the sky map a bolometric correction to $\rm UL(\Omega)$ considering the spectral template that produces the largest, and hence most conservative, flux upper limit in the same energy band.
From the posterior samples of GW230529 we extract the distribution of RA, Dec, luminosity distance and inclination angle. Among these parameters, only the inclination angle is a free parameter in our modeling, while we do not make any further inference about the sky position and the distance. Calling $\Vec{\theta}$ the set of intrinsic parameters of the jet structure, the likelihood of the model is defined as
\begin{equation}
\label{like}
    \mathcal{L}(\Vec{\theta},\theta_v)=P_{ND}(\Vec{\theta},\theta_v),
\end{equation}
corresponding to the probability of non-detection.

\begin{table*}[]
\centering
    \begin{tabular}{|l|c|c|c|c|c|c|cc}
        \hline
        &\multicolumn{3}{c|}{Prior}&\multicolumn{3}{c|}{Posterior} \\
        \hline
         \multirow{2}{*}{jet structure} & $L_0$  & $\theta_c$ &  $\theta_v$ &  $L_0$  & $\theta_c$ &  $\theta_v$ \\
         & erg s$^{-1}$  & deg &  deg &  erg s$^{-1}$ & deg &  deg \\
         \hline
\textit{top--hat} & $10^{50}$ $(10^{54})$  & 15.1 (27.0) &  $40.0^{+19.1}_{-18.2}$ &  $3.4\times 10^{48}$ $(4.7\times 10^{53})$  & 11.3 (27.3) &  $43.6^{+18.2}_{-19.1}$ \\

\textit{Gaussian} & $10^{50}$ $(10^{54})$  & 15.1 (27.0) &  $40.0^{+19.1}_{-18.2}$ &  $2.1\times 10^{47}$ $(2.3\times 10^{50})$  & 12.5 (28.8) &  $40.9^{+19.1}_{-18.2}$ \\

\textit{power law}, $s=2$& $10^{50}$ $(10^{54})$  & 15.1 (27.0) &  $40.0^{+19.1}_{-18.2}$ &  $1.7\times 10^{47}$ $(1.7\times 10^{49})$  & 14.0 (28.5) &  $40.9^{+18.2}_{-18.2}$ \\

\textit{power law}, $s=4$ & $10^{50}$ $(10^{54})$  & 15.1 (27.0) &  $40.0^{+19.1}_{-18.2}$ &  $2.7\times 10^{47}$ $(1.1\times 10^{50})$  & 12.5 (28.5) &  $40.9^{+19.1}_{-18.2}$ \\   
        \hline

    \end{tabular}
    
    \label{tab:ul}

\centering
    \begin{tabular}{|c|c|c|c|c|c|c|c|}

        \multicolumn{8}{c}{}\\ 
        \multicolumn{8}{c}{\textit{power law}, free s}\\
        \hline
        \multicolumn{4}{|c|}{Prior}&\multicolumn{4}{c|}{Posterior} \\
        \hline
           $L_0$  & $\theta_c$ &  $\theta_v$ & s&  $L_0$ & $\theta_c$ &  $\theta_v$  &s \\
         erg s$^{-1}$  & deg &  deg & - &  erg s$^{-1}$ & deg &  deg & -\\
         \hline
         
            $10^{50}$ $(10^{54})$  & 15.1 (27.0) &  $40.0^{+19.1}_{-18.2}$  & 1.9 (5.5) &  $1.2\times 10^{48}$ $(7.1\times 10^{51})$  & 12.5 (26.0) &  $41.8^{+18.1}_{-17.7}$ & 2.2 (6.1) \\ \hline
           \multicolumn{8}{c}{}\\ 
           
            \multicolumn{8}{c}{\textit{top--hat}+\textit{isotropic} component}\\
            \hline
        \multicolumn{4}{|c|}{Prior}&\multicolumn{4}{c|}{Posterior} \\ \hline
        $L_0$  & $\theta_c$ &  $\theta_v$ & $L_{\rm off}$&  $L_0$ & $\theta_c$ &  $\theta_v$  &$L_{\rm off}$ \\
         erg s$^{-1}$  & deg &  deg & erg s$^{-1}$ &  erg s$^{-1}$ & deg &  deg & erg s$^{-1}$\\
         \hline           
        
         $10^{50}$ $(10^{54})$  & 15.1 (27.0) &  $40.0^{+19.1}_{-18.2}$  & $10^{45}$ $(10^{49})$ &  $5.9\times 10^{49}$ $(9.1\times 10^{53})$  & 14.3 (26.6) &  $40.7^{+17.7}_{-16.3}$ & $1.9\times 10^{44}$ $(5.2\times 10^{47})$ \\          
        \hline
    \end{tabular}
    
    \caption{The tables show the limits on the model parameters, comparing the percentiles of the prior and posterior distributions, for different jet structures. We compare the percentiles of the prior and posterior distributions. The upper limits on $L_0$ and $\theta_c$ are reported as 50$^{\rm th}$ (90$^{\rm th}$) percentiles. The credible interval of $\theta_v$ is at 1$sigma$. Since the posterior $P(s)$ is monotonically increasing for larger values of $s$, we report lower limits as 10$^{\rm th}$ (50$^{\rm th}$) percentiles.}
    \label{tab:sfree}
\end{table*}

For a fixed combination of $(\Vec{\theta},\theta_v)$, the non-detection probability depends also on the distance and on the sky coordinates, since the sensitivity of the $\gamma-$ray monitors changes as a function of the sky position. The non-detection probability can be explicated as
\begin{gather}
    P_{ND}(\Vec{\theta},\theta_v)= \notag \\
    \int P(F<\text{UL}(\Omega)) P_{GW}(\Omega, D_L|\theta_v)  d \Omega d D_L,
\end{gather}
where $P(F<\text{UL}(\Omega))$ is the probability that, having fixed the parameters $(\Vec{\theta},\Omega, D_L,\theta_v)$, the observed flux is below the upper limit $\text{UL}(\Omega)$, while  $\Omega$ is the sky position (RA, Dec), $D_L$ is the luminosity distance, $P_{GW}(\Omega, D_L|\theta_v)$ is the corresponding conditional probability distributions extracted from the GW posteriors, which in turn depend on $\theta_v$. The function $\mathcal{L}(\Vec{\theta},\theta_v)$ has been evaluated injecting sources according to the distributions $P_{GW}(\Omega,D_L|\theta_v)$. The non-detection probability is computed by counting the fraction of injected sources that produced an observed flux in the 1 keV--10 MeV band lower than the upper limit $\rm UL(\Omega)$.

According to Bayes theorem, we computed the posterior distribution as
\begin{equation}
    \mathcal{P}(\Vec{\theta},\theta_v)\propto \mathcal{L}(\Vec{\theta},\theta_v) \prod_j \pi(\theta_j) ,
\end{equation}
where $\pi(\theta_j)$ are the respective priors. As mentioned before $\pi(\theta_v)$ corresponds to the GW posterior of the viewing angle. For the model parameters $L_0$, $L_{\rm off}$, $\theta_c$ and $s$ we consider uniform priors
\begin{equation}
    \pi[\log(L_0)]=
     \begin{cases}
\rm const, & 45 <\log[L_0(\rm erg/s)] <55 \\
0, & \rm otherwise
\end{cases}
\end{equation}

\begin{equation}
    \pi[\log(L_{\rm off})]=
     \begin{cases}
\rm const, & 40 <\log[L_{\rm off}(\rm erg/s)] <50 \\
0, & \rm otherwise
\end{cases}
\end{equation}

\begin{equation}
    \pi(\theta_c)=
     \begin{cases}
\rm const, & 0.1 \rm\, deg <\theta_c <30 \rm \,deg \\
0, & \rm otherwise
\end{cases}
\end{equation}

\begin{equation}
    \pi(s)=
     \begin{cases}
\rm const, & 1  <s <10  \\
0, & \rm otherwise
\end{cases}.
\end{equation}

The choice of $\pi[\log(L_0)]$ fully covers the range of possible luminosities a GRB could have, from the sub-luminous regime to the typical values of cosmological GRBs (e.g., \citealt{2015ApJ...815..102F}). The boundaries in the prior $\pi(\theta_c)$ are compatible with previous values of half-jet opening angles derived in the literature \citep{2015ApJ...815..102F,2016ApJ...827..102T,2019ApJ...883...48L,ruoco}. The prior of $L_{\rm off}$, instead, is bounded to lower values, since the off-axis emission of a GRB is expected to be less Doppler boosted and, therefore, sub-luminous with respect to the on-axis component \citep{2002MNRAS.337.1349R,2017ApJ...834...28N, 2017ApJ...848L...6L,2017MNRAS.471.1652L,2018MNRAS.473..576G,2023MNRAS.520.1111H}.
The distribution $\pi(\theta_v)$, corresponding to the GW posterior $P_{GW}(\theta_v)$, is close to being symmetric around $\pi/2$. Moreover, the jet emission is assumed to be symmetric with respect to the binary orbital plane, namely $L(\theta_v)=L(\pi/2 -\theta_v)$, therefore, hereafter we consider the inclination angle $\theta_v$ only in the range $[0,\pi/2]$. This is obtained by creating a prior 
\begin{equation}
    \pi(\theta_v)=
         \begin{cases}
P_{GW}(\theta_v)+P_{GW}(\frac{\pi}{2}-\theta_v), & 0  <\theta_v <\frac{\pi}{2}  \\
0, & \rm otherwise
\end{cases},
\end{equation}
and then re-normalized to have $\int \pi(\theta_v)=1$. For the derivation of the posterior distribution, we adopt the following approaches.
\begin{enumerate}

    \item We perform a numerical evaluation of the likelihood for all the models with three or fewer free parameters, namely the \textit{top--hat}, the \textit{Gaussian}, the \textit{power law} with fixed $s$ and the \textit{isotropic} model. Each model parameter is sampled in a linearly spaced bin of 30 elements.
    \item The \texttt{emcee} sampler \citep{2013PASP..125..306F} is used for all the models with more than three free parameters, namely the \textit{power-law} with free $s$ and the \textit{top--hat}+\textit{isotropic} component. The convergence of the chain is controlled imposing that the auto-correlation time is larger than 50 times the number of steps.
    
\end{enumerate}
\section{Results and discussion} \label{sec:results}

\begin{figure}
    \centering
    \includegraphics[width=1.0\columnwidth]{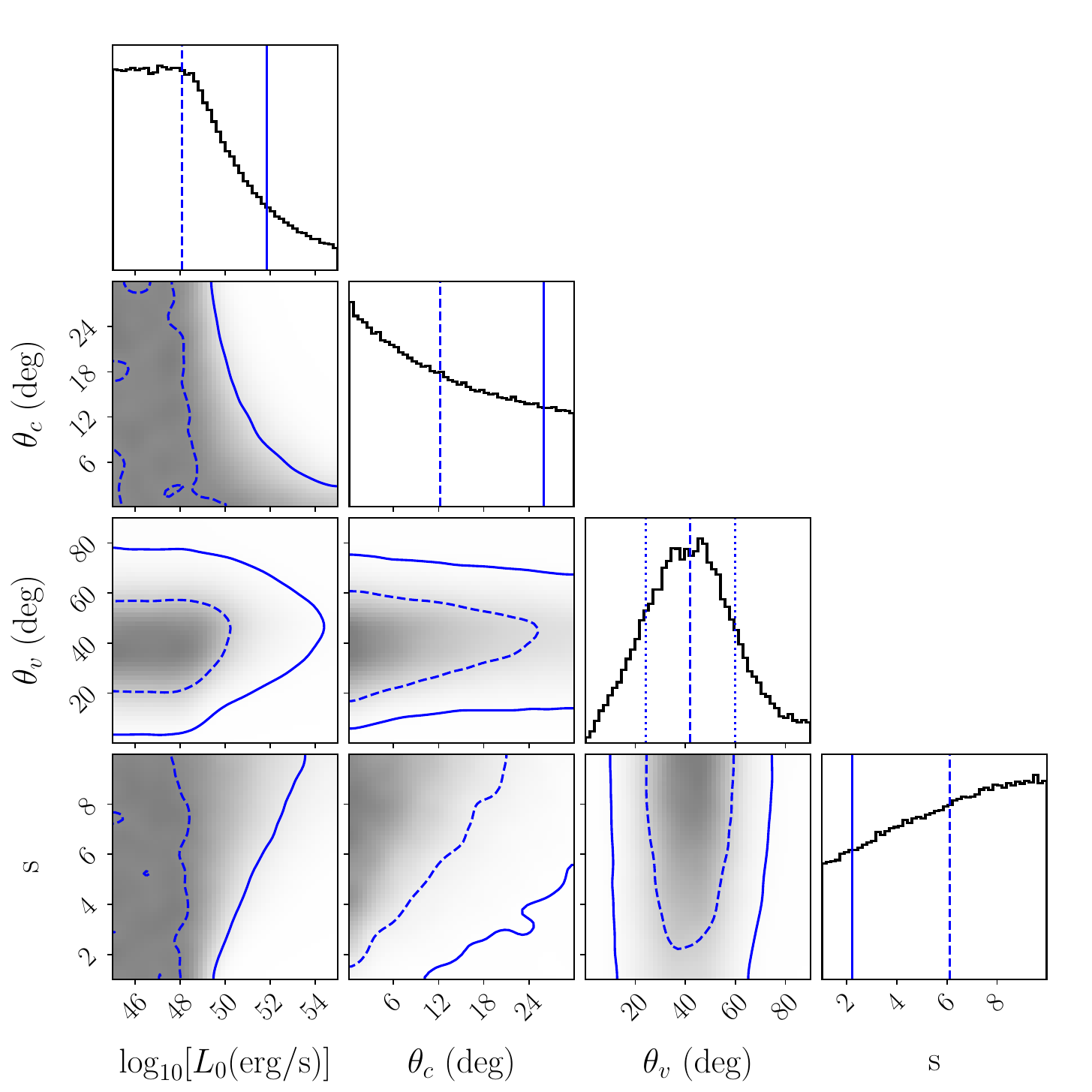}
    \includegraphics[width=1.0\columnwidth]{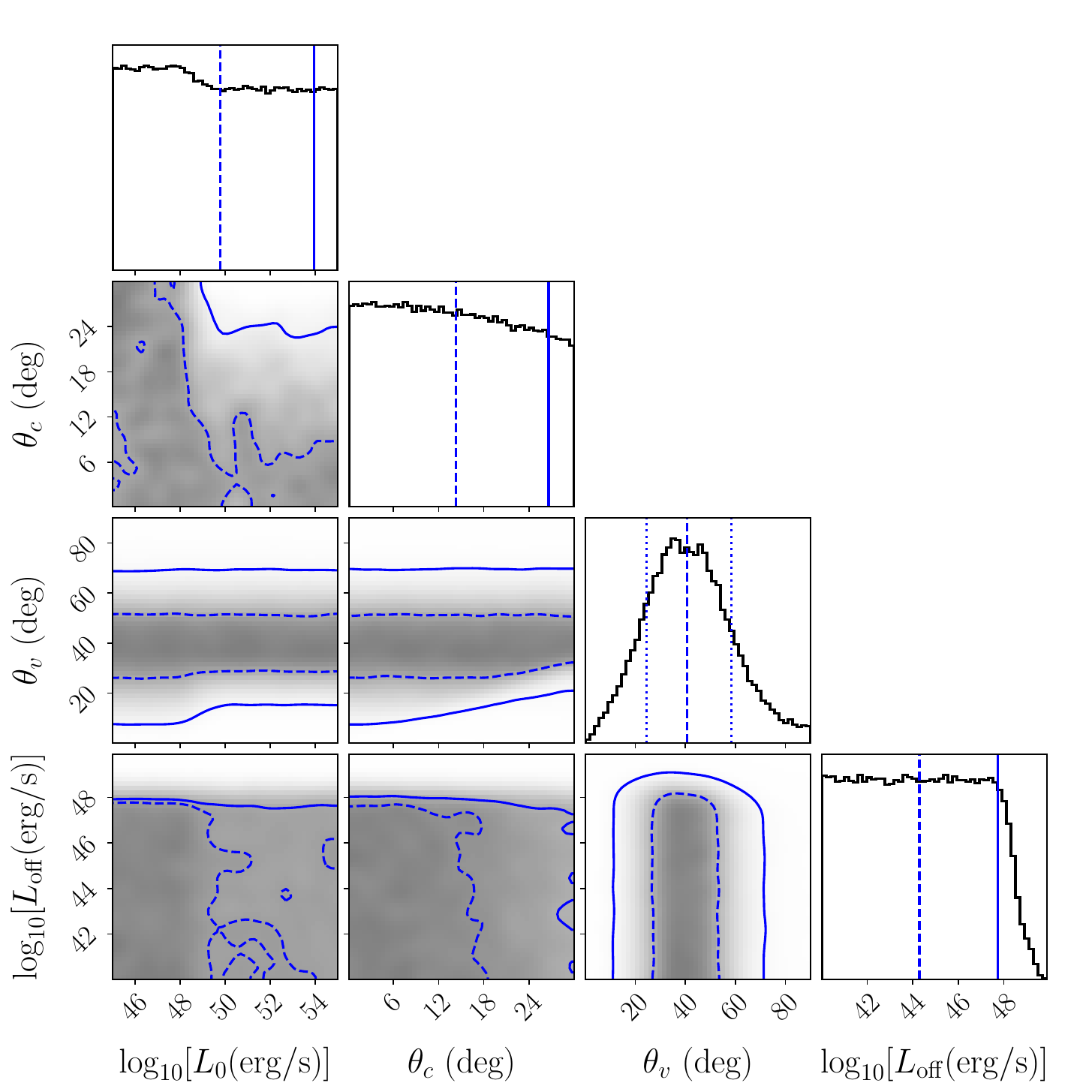}
    \caption{Same as Fig.~\ref{corner_tophat}, but for a \textit{power law} model with $s$ as free parameter (top panel) and the \textit{top--hat}+\textit{isotropic} component (bottom panel). For the panel of the posterior distribution of $s$, the solid line reports the 10$^{\rm th}$ percentile.}
    \label{fig:frees+twocomp}
\end{figure}

\begin{figure}
    \centering
    \includegraphics[width=1.0\columnwidth]{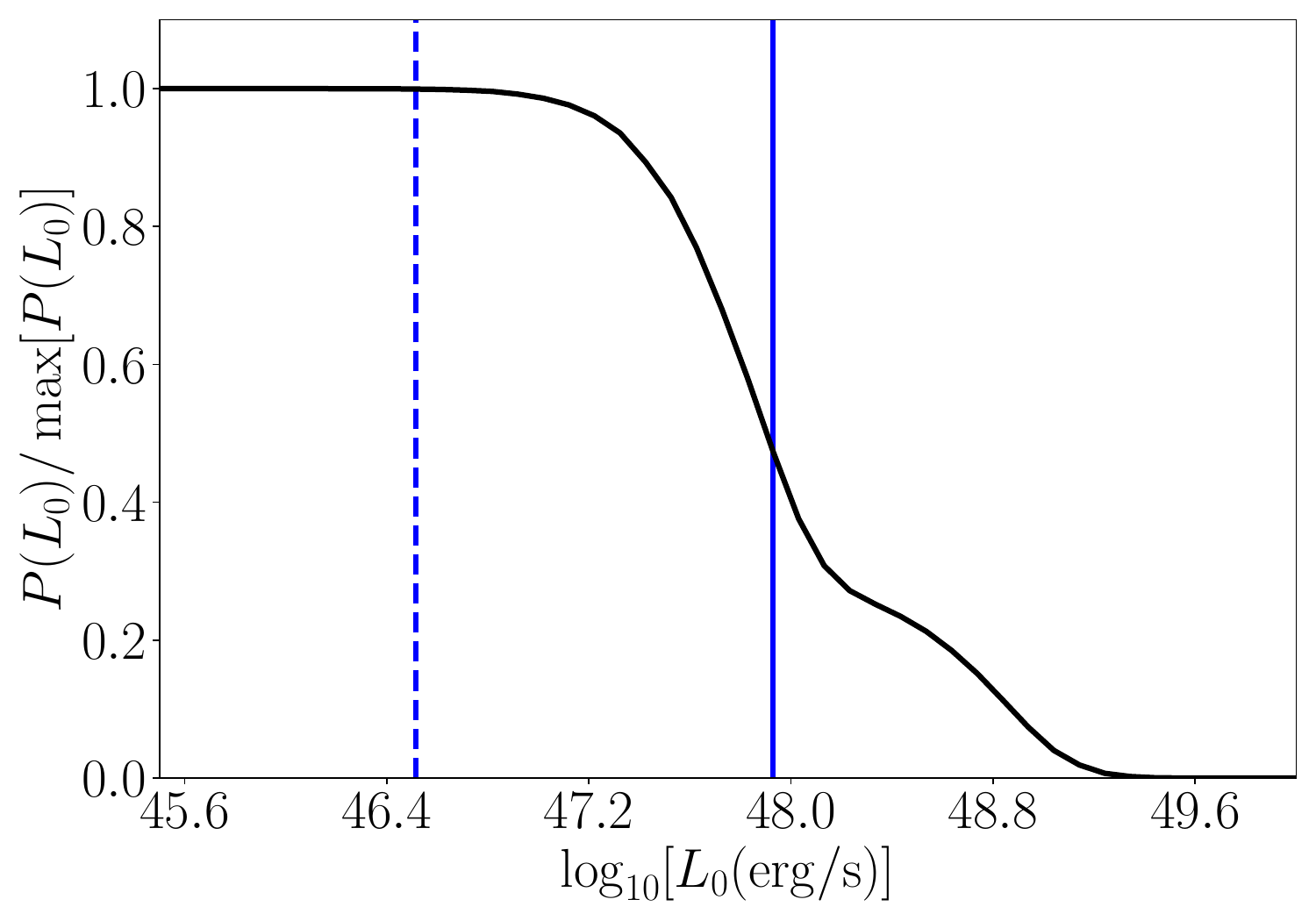}
    \caption{Posterior distribution of the apparent isotropic luminosity for the \textit{isotropic} model. The solid and dashed vertical lines are the 90$^{\rm th}$ and 50$^{\rm th}$ percentiles of the distribution, respectively.}
    \label{fig:iso}
\end{figure}

\begin{figure*}
    \centering
    \includegraphics[width=1.0\textwidth]{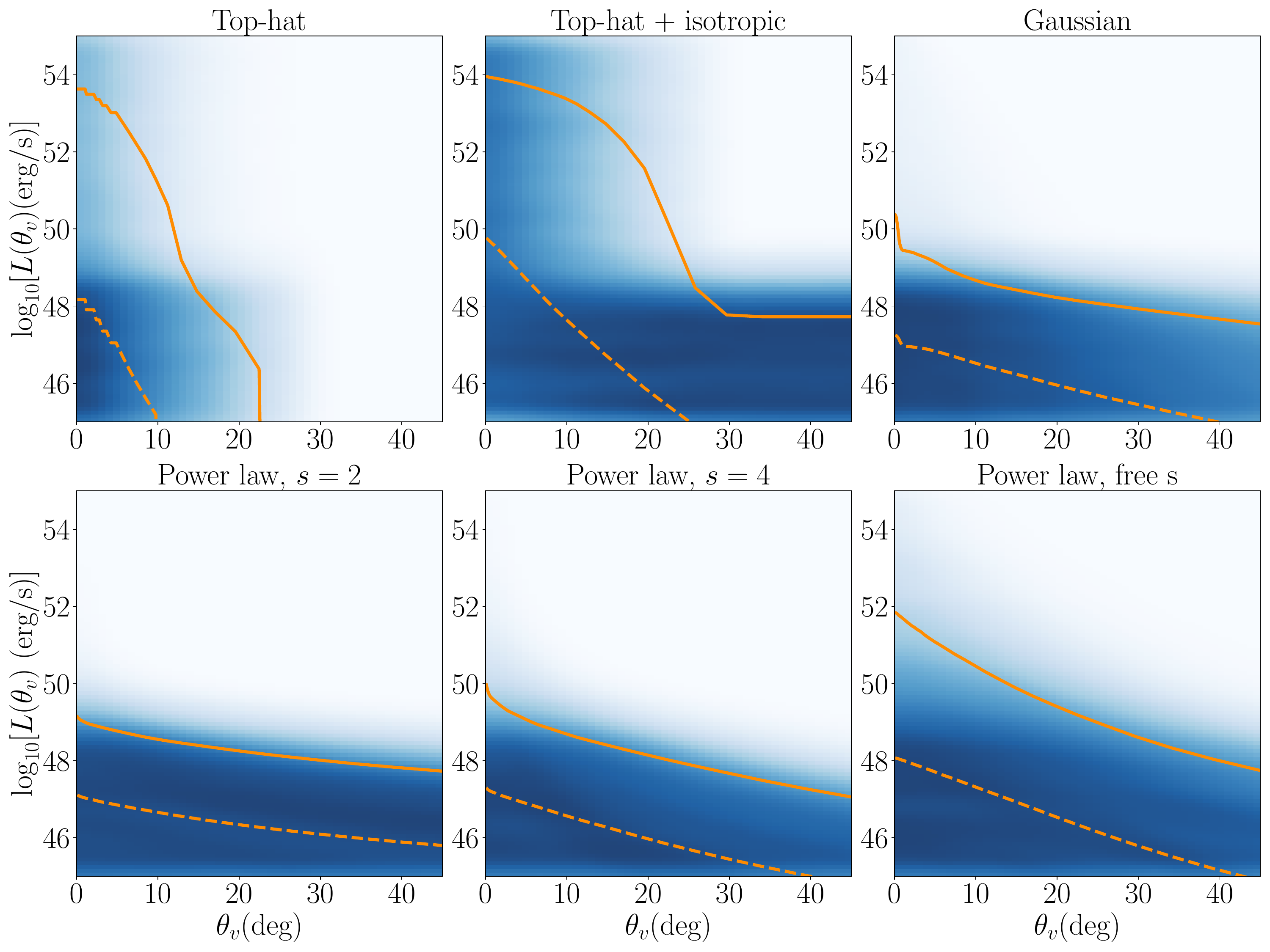}
    \caption{Posterior distribution of the luminosity profile $L(\theta_v)$, assuming different angular structures. The color density is produced extracting randomly $L_0$, $\theta_c$, $s$ and $L_{\rm off}$ from the relative posterior samples, which are then given as input to compute the luminosity profile $L(\theta_v)$ for each extraction. For each subplot, the solid (dashed) lines indicate the 90$\%$ (50$\%$) upper limits.}
    \label{l_theta_post}
\end{figure*}

The corner plot of the posterior distribution of $L_0$, $\theta_c$ and $\theta_v$ is shown in Fig.~\ref{corner_tophat} for the \textit{top--hat} structure, in Fig.~\ref{corner_gauss} for the \textit{Gaussian}, and in Fig.~\ref{corner_pl} for the \textit{power law} with $s=2$ and $s=4$, respectively. Darker regions inside each panel indicate where the posterior probability is higher. Therefore, the excluded regions are outside the gray areas.  Focusing on the \textit{top--hat} plot, we note the following:
\begin{enumerate}
    \item Analyzing the $\theta_c$--$L_0$ posterior panel, we can derive that a jet configuration with $L_0\gtrsim 10^{48}$ erg s$^{-1}$ and $\theta_c\gtrsim 15$ deg is excluded at 90$\%$ credibility.
    
    \item Analyzing the $\theta_v-L_0$ posterior panel, we can conclude that, if the jet had an on-axis luminosity of $L_0\gtrsim 10^{50}$ erg s$^{-1}$ our analysis implies that the viewing angle could not be smaller than $\sim$ 15 deg at $90\%$ credibility, independently of the opening angle of the jet and of the assumption about the jet structure.
    
    \item From the comparison in the central panel between the prior $\pi(\theta_v)$  and the posterior $P(\theta_v)$, our analysis shows an overall marginal preference for larger values of $\theta_v$. 
\end{enumerate}
Very similar behaviors are present in the corner plots of the other jet structures, with the major difference that the $90\%$ exclusion region in the $\theta_c$--$L_0$ posterior panel extends to lower values of $\theta_c$. Such a behavior is expected, since all the jet profiles that differ from a \textit{top--hat} have emission also outside $\theta_c$, implying more stringent constraints on this parameter.

Considering the posterior distribution of $L_0$, especially in the \textit{top--hat} scenario, it is evident that $P(L_0)$ flattens and stays constant for values above $\sim 10^{50}$ erg s$^{-1}$. This can be explained as follows. Let us consider two values of $L_0$ and the corresponding likelihoods $P(L_{0,a})$ and $P(L_{0,b})$. If both $L_{0,a}$ and $L_{0,b}$ are large enough, then whenever $\theta_v<\theta_c$ the predicted flux is above the derived upper limit. Namely,
\begin{equation}
    P(F<\text{UL}(\Omega))\sim \Theta(\theta_v-\theta_c),
\end{equation}
where $\Theta(x)$ is a step function. Therefore, the likelihood becomes independent of $L_0$ and saturates to a constant value equal to
\begin{gather}
    \mathcal{L}(L_0,\theta_c,\theta_v)= \\
    \int \Theta(\theta_v-\theta_c) P_{GW}(\Omega,D_L|\theta_v) d \Omega d D_L.
\end{gather}

In Fig.~\ref{fig:frees+twocomp} we show the corner plot for the \textit{power law} structure with free $s$ (top panel) and the \textit{top--hat}+\textit{isotropic} component (bottom panel). For the \textit{power law} structure, as expected, the posterior distribution of $s$ peaks at high values, giving preference to a jet whose off-axis emission sharply drops outside the core. From the $\theta_c$--$L_0$ correlation plot, we can see that for a representative value of $\theta_c\sim 10$ deg (typical of standard GRBs) our analysis imposes an upper limit on the on-axis isotropic luminosity of $\sim 10^{51}$ erg s$^{-1}$ (90$\%$ credibility). For the \textit{top--hat}+\textit{isotropic} component structure, the $L_0$ and $\theta_c$ parameters are more poorly constrained with respect to the other structures. On the other hand, the off-axis luminosity $L_{\rm off}$ is constrained to be $\lesssim 10^{48}$ erg s$^{-1}$ (90$\%$ credibility).

The plot in Fig.~\ref{fig:iso} shows the posterior distribution of the apparent luminosity for a hypothetical sub-luminous isotropic emission. Since the flux upper limits for this model are computed using a temporal bin of 16.384 s, $L_0$ has to be the average luminosity on the same timescale. The 90$\%$ credible upper limit on $L_0$ is $8.5\times 10^{47}$ erg s$^{-1}$.

In Fig.~\ref{l_theta_post} we show the posterior probability map of the luminosity profile $L(\theta_v)$, for each jet structure. The map is obtained by extracting the jet structure parameters ($L_0$, $\theta_c$, $s$, $L_{\rm off}$) from the respective posterior distributions, for each jet structure considered in this work. For a given fixed value of $\theta_v$ we report the $50\%$ and $90\%$ upper limits of $L(\theta_v)$, indicated with a dashed and solid line, respectively. As it can be noticed from this figure, apart from the \textit{top--hat} structure, some values of $L_0$ and $\theta_c$ produce an $L(\theta_v)$ unphysically bright at large viewing angles. This effect is driven by the choice of the priors of $L_0$ and $\theta_c$, which are not physically informed by any assumption about the maximum luminosity of a jet viewed strongly off-axis. In order to have such large luminosities off-axis, the jet should move relativistically not only along its axis but also along the line of sight, namely outside the jet core. This is likely not physical since it would by far exceed the energy budget of the jet, limited by the amount of mass accreted around the central engine. Therefore, for viewing angles $\theta_v\gg10$ deg (typical opening angle of GRB jets) the upper limits on $L(\theta_v)$ shown in Fig.~\ref{l_theta_post} have to be taken with caution. However, we underline that, since we provide only upper limits on $L(\theta_v)$, our choice of the prior possibly overestimates the limits, making them more conservative and, therefore, not impacting the robustness of our conclusions.

A summary of all the properties of the posterior distributions are reported in Tab.~\ref{tab:sfree}. The tables compare the percentiles of the priors and the posteriors of each parameter. For all the posteriors that show a decreasing trend we report the upper limits as the $50^{\rm th}$ and $90^{\rm th}$ percentiles. In the case of the parameter $s$, since the posterior favors larger values, we report the lower limits as the $10^{\rm th}$ and $50^{\rm th}$ percentiles. The posterior of $\theta_v$ is well peaked, as the prior, and we report the 1$sigma$ credible interval ($16^{\rm th}$ and $84^{\rm th}$ percentiles) and the median value.
\begin{figure}
    \centering
    \includegraphics[width=1.0\columnwidth]{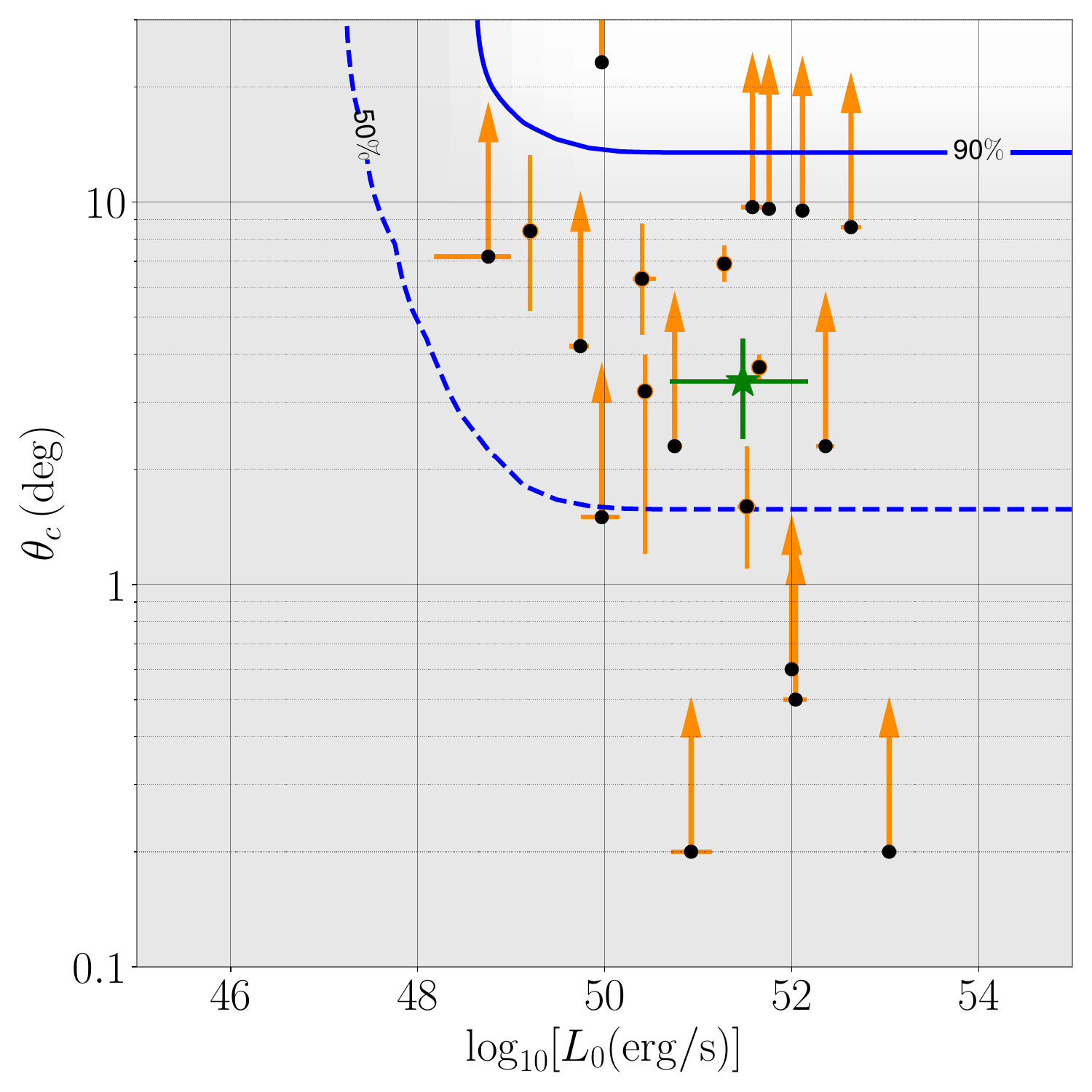}
    \caption{$L_0$--$\theta_c$ posterior probability (taken from the lower left panel of Fig.~\ref{corner_tophat}) compared with estimates from \cite{ruoco} of isotropic luminosities and opening angles of a sample of short GRBs. The green point is relative to GW170817, for which the on-axis isotropic luminosity is taken from \cite{2019A&A...628A..18S} and the opening angle from \cite{2019Sci...363..968G}. The solid and dashed lines are 90$\%$ and 50$\%$ exclusion regions derived from the present analysis, in the assumption of a \textit{top--hat} jet structure.}
    \label{fig:ruoco}
\end{figure}
\begin{figure}
    \centering
    \includegraphics[width=1.0\columnwidth]{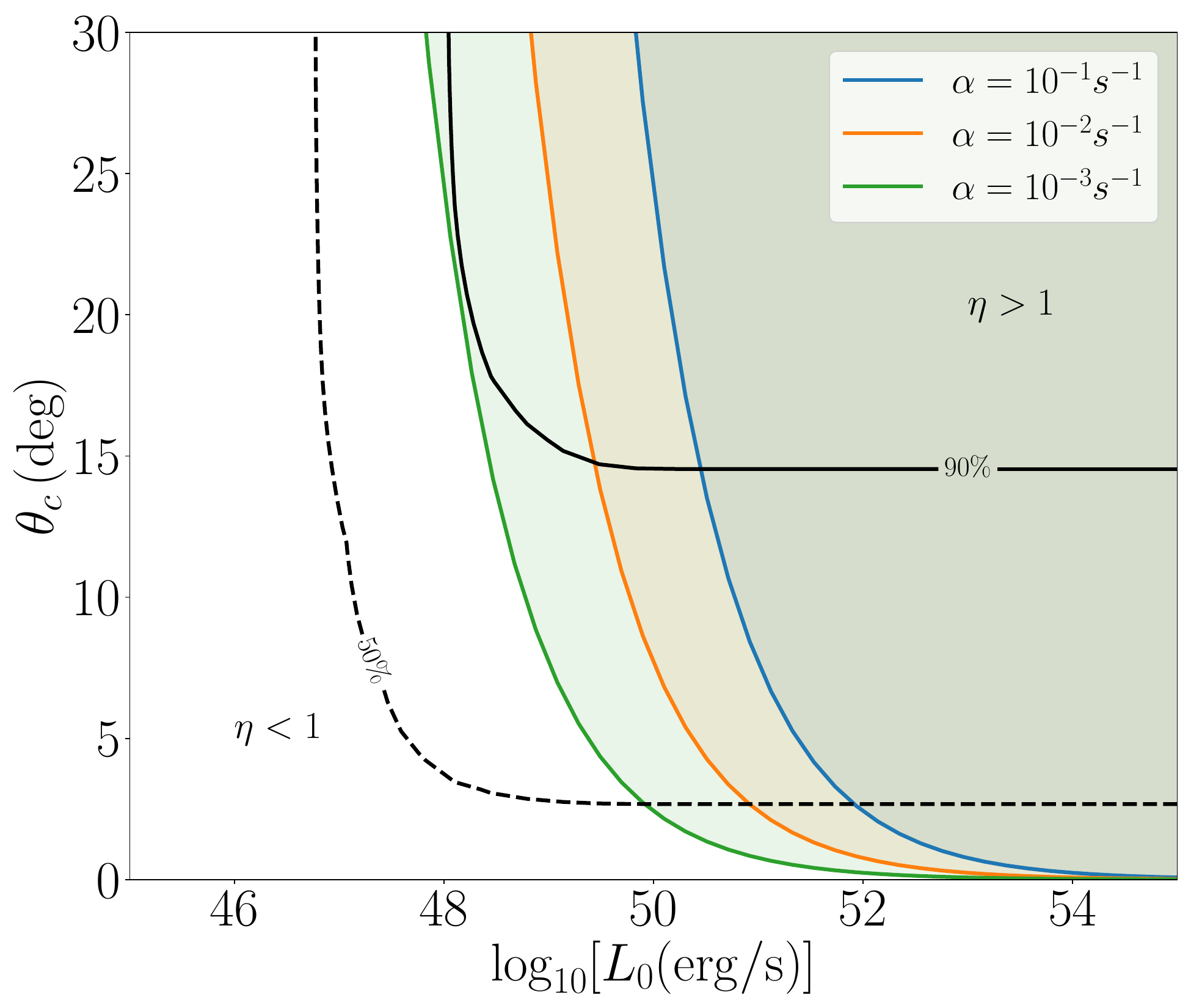}
    \caption{Regions of the $L_0$--$\theta_c$ plane excluded by the condition $M_{ej}<0.052 M_{\odot}$. The exclusion lines are computed for different values of $\alpha$, defined in the text. The dashed and solid black lines are the $50\%$ and $90\%$ exclusion regions reported from Fig.~\ref{corner_tophat}.}
    \label{fig:forb}
\end{figure}

Since the GW merger considered in this work can potentially power a short GRB, it is useful to compare the constraints obtained here with the typical jet luminosities and opening angles collected in the literature. As a reference, we consider the short GRBs analyzed by \cite{ruoco}. The authors combine observations of the prompt and afterglow phases of short GRBs, deriving the energetics of the jet and the opening angle of the jet from the detection of a jet-break in the afterglow lightcurve. In particular, we consider the values of the isotropic equivalent energy $E_{\rm iso}$ and the opening angle $\theta_j$ reported in Tab.~4 of \cite{ruoco}. In order to have an estimate of the isotropic equivalent luminosity we compute $(1+z)E_{\rm iso}/t_{90}$, $t_{90}$ being the timescale over which we receive 90$\%$ of the observed photons of the burst. Since the isotropic equivalent luminosity corresponds to the definition of our $L_0$ in the case of a \textit{top--hat} jet, we compare in Fig.~\ref{fig:ruoco} the $L_{\rm iso}$--$\theta_j$ values derived from \cite{ruoco} with the 2D posterior distribution of $L_0$--$\theta_c$. We indicated with an arrow the GRBs for which we have only a lower limit of the opening angle $\theta_j$. Comparing the 90$\%$ excluded region with the location of the $L_{\rm iso}$--$\theta_j$ points, we do not exclude that GW230529 produced a typical short GRB-like jet.

The value of the opening angle of the jet can be compared with the results of \cite{Sarin2022PhRvD}, where the authors combine the estimated rate of BNS and NSBH mergers with the rate of observed short GRBs, showing that the average opening angle of jets launched by NSBH mergers is in the range $15.3^{+3.0}_{-3.3}$ deg (1$\sigma$ credibility), under the assumption of a top-hat jet structure. With the caveat that this range could be dependent on the assumptions on the NSBH population and the modelling of the jet emission, the authors find a value compatible with the exclusion regions derived in this work. Considering different EoS, \cite{Zhu2023arXiv} estimate the amount of mass ejected by GW230529, including the fraction that goes into an accretion disk. Assuming a Blandford–Znajek jet with a Gaussian structure, they derive a probability of detecting a GRB-like emission $<5\%$, in line with the non-detection reported in this paper.

Our analysis provides limits on the maximum on-axis isotropic luminosity and the jet opening angle. Therefore, it is possible to infer a limit on the beaming corrected luminosity, defined as $L_b=L_{0}(1-\cos \theta_c)$. Since the GW analysis reports an upper limit on the amount of ejected mass $M_{ej}<0.052 M_{\odot}$, this represents also a limit on the maximum amount of mass bound to the central object that can potentially power the GRB jet. If we define $\eta$ as the efficiency of conversion from the rest mass energy of the bound disk mass to the beaming corrected energy of the jet,
we can write
\begin{equation}
    L_{b} = \eta \frac{M_{acc} c^2}{t_{acc}}=\eta \dot{M}_{acc} c^2,
\end{equation}
where $M_{acc}$ is the fraction of the ejected mass which falls back and accretes around the central BH, while $t_{acc}$ is the accretion timescale of the disk.
Using the limits $L_0\lesssim 10^{48}$ erg s$^{-1}$ and $\theta_c\lesssim 15$ deg found for the \textit{top--hat} structure, we derive a limit on $\eta$ corresponding to
\begin{equation}
    \eta \lesssim 1.9 \times 10^{-3} \left(\frac{M_{acc}}{0.01 M_{\odot}} \right)^{-1}  \left(\frac{t_{acc}}{1\text{ s}} \right).
\end{equation}

\begin{figure}
    \centering
    \includegraphics[width=1.0\columnwidth]{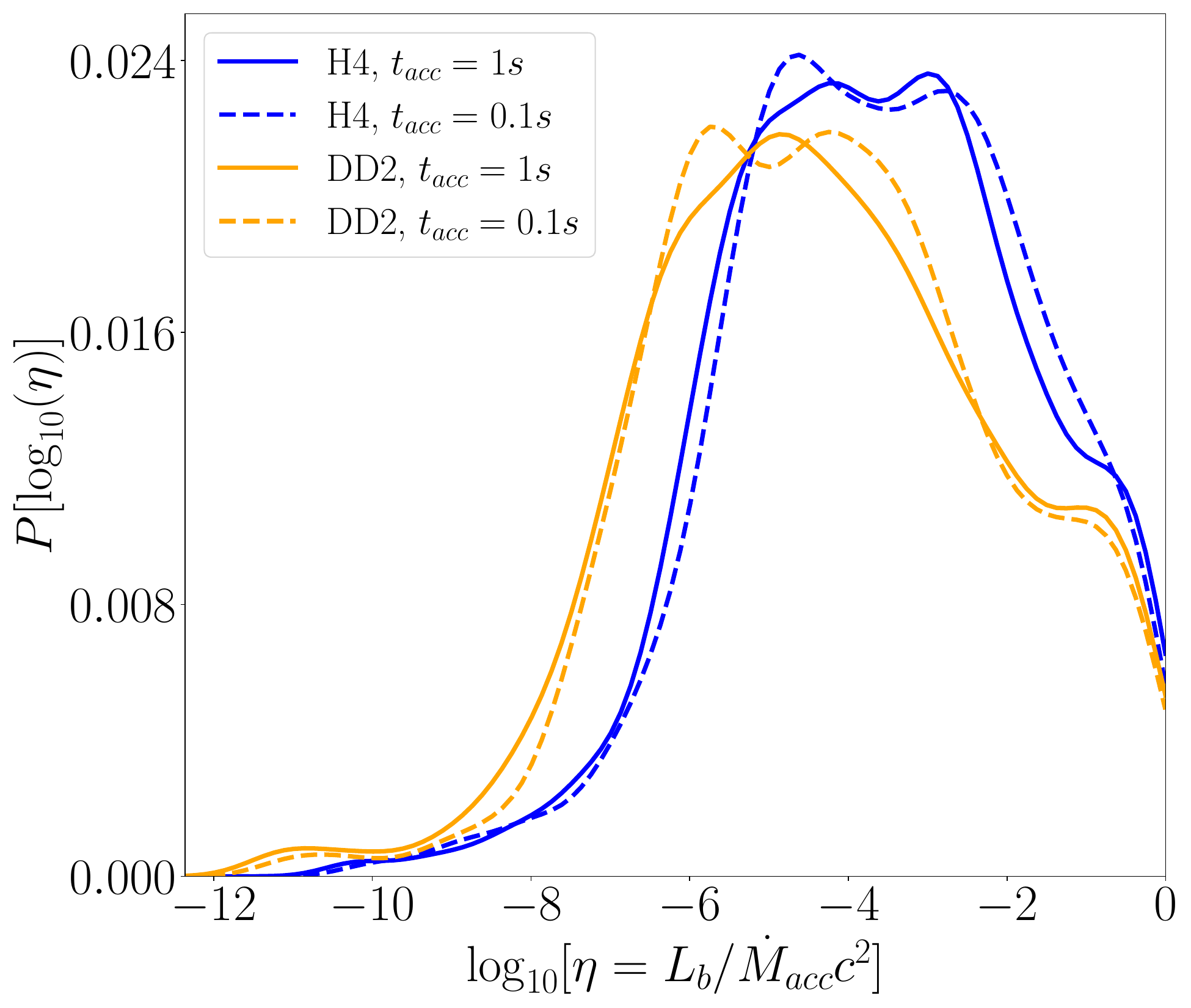}
    \caption{Posterior distribution of the efficiency factor $\eta$, which is the fraction of accretion power which is converted in the beaming-corrected luminosity of the jet. The posterior is produced for the EoS \texttt{H4} and \texttt{DD2}, assuming an accretion time scale of 1 s and 0.1 s.}
    \label{fig:eta}
\end{figure}

Since the conversion efficiency $\eta$ cannot be larger than one, we can exploit the upper limit on $M_{ej}$ derived from the GW analysis to obtain a limit on the maximum allowed opening angle of the jet, for a given value of $L_0$. Expressing $M_{acc}=\epsilon M_{ej}$ and including all our uncertainties in a parameter $\alpha=\epsilon/t_{acc}$, we can impose that $\eta<1$ and $M_{ej}<0.052M_{\odot}$,  obtaining:
\begin{equation}
    \theta_c<\arccos \left( 1- \alpha \frac{0.052 M_{\odot}c^2}{L_0} \right).
\end{equation}
In Fig.~\ref{fig:forb} we report the regions of the $L_0$--$\theta_c$ plane that are not allowed by the condition $\eta<1$ and $M_{ej}<0.052M_{\odot}$, for different values of $\alpha$, over-plotted to the $50\%$ and $90\%$ exclusion regions derived previously for the \textit{top-hat} jet. The figure shows that the smaller the fraction of ejected mass that goes into an accretion disk, the larger is the excluded region in the $L_0$--$\theta_c$ plane. The values of $L_0$--$\theta_c$ excluded by the condition $M_{ej}<0.052M_{\odot}$ are in agreement with the region excluded by our analysis, showing that the non-detection of $\gamma-$rays is compatible with the limited amount of ejected mass that can be channeled to power a GRB jet.

In order to infer more informative constraints on $\eta$, we consider the work performed by \cite{Chandra2024}. The authors use a set of population synthesis models to derive the mass distribution of the two components of GW230529. In addition, adopting different EoS, they estimate the mass of the remnant BH, as well as the amount of ejecta mass. The latter is calculated following the fitting formulae derived by \cite{2020PhRvD.101j3002K} and \cite{2021ApJ...922..269R}. Here we consider only the two EoS \texttt{H4} \citep{PhysRevLett.67.2414,PhysRevD.73.024021,PhysRevD.79.124032} and \texttt{DD2} \citep{PhysRevC.81.015803,HEMPEL2010210} that produce ejecta mass larger than zero. From the ejecta mass, for each EoS we extract the posterior distribution of the bound disk mass. The final distribution of $\eta$ is obtained randomly extracting $L_0$ and $\theta_c$ from the \textit{top--hat} posterior samples, combined with the distribution of the bound disk mass. The result is shown in Fig.~\ref{fig:eta}, for the two EoS \texttt{H4} and \texttt{DD2}, and assuming $t_{acc}=0.1$ s and $t_{acc}=1$ s, in the typical range expected for NSBH mergers \citep{2018IJMPD..2742004C}. The 90$\%$ credibility posterior range is $\log_{10}\eta = -3.6_{-2.4}^{+2.6}$ ($\log_{10}\eta = -3.6_{-2.3}^{+2.5}$) for the \texttt{H4} EoS and $t_{acc}=1$ s ($t_{acc}=0.1$ s), and $\log_{10}\eta = -4.4_{-2.5}^{+3.2}$ ($\log_{10}\eta = -4.3_{-2.5}^{+3.1}$) for the \texttt{DD2} EoS and $t_{acc}=1$ s ($t_{acc}=0.1$ s).

\section{Conclusions}

GW230529 is the first NSBH candidate with a confirmed primary component in the lower mass-gap. Being the NSBH event with the most symmetric mass ratio detected up to date, this event is particularly promising for the production of electromagnetic emission in the form of a relativistic jet and/or a kilonova. No EM counterpart associated with this event has been reported so far \citep{2023GCN.33890....1S,2023GCN.33892....1L,2023GCN.33893....1S,2023GCN.33894....1L,2023GCN.33896....1W,2023GCN.33897....1S,2023GCN.33900....1K,2023GCN.33980....1I}. In the $\gamma-$ray domain, the combined presence of \textit{Swift} and \textit{Fermi} telescopes ensured the full coverage of the sky at the time of the GW trigger. This enables us to derive the most stringent constraints on the possible GRB-like emission from the GW candidate. Combining the posterior distribution of sky position, luminosity distance, and inclination angle derived from the GW parameter estimation, we use the \textit{Swift}--BAT and \textit{Fermi}--GBM flux upper limits to derive the allowed values of the luminosity and opening angle of the putative jet. The analysis focused on multiple jet structures, including a top-hat, a Gaussian, a power law, as also the inclusion of a possible isotropic component outside the jet core. The most conservative constraints are obtained for the \textit{top--hat} configuration, for which there is a non-negligible probability that the jet, even if bright, was strongly off-axis and, therefore, non-detectable. Considering the on-axis isotropic luminosity $L_0$ of a top-hat jet (in the rest frame energy band 1 keV--10 MeV), we conclude that the non-detection of a $\gamma-$ray emission implies the absence of a jet with a $L_0\gtrsim 10^{48}$ erg s$^{-1}$ and an opening angle $\gtrsim$ 15 deg. These limits become stronger as soon as we admit the presence of an off-axis emission. Comparing the derived limits on the isotropic luminosity and opening angle with the typical values from short GRBs observed at cosmological distances, we cannot exclude that a system like GW230529 can potentially power a standard merger-driven GRB. 

The absence of electromagnetic counterparts to GW230529, as constrained by the joint Fermi-Swift search reported here, opens up the possibility that this event is a primordial black hole merger \citep{2024arXiv240405691H}, similar to that of GW190425 and GW190814 \citep{Clesse:2020ghq}. In fact, there is no information from the GW analysis about the tidal deformability of the secondary component. Therefore, the possibility that the secondary was a primordial BH cannot be totally excluded. The two component masses of GW230529 are in the mass-gap around 2$M_{\odot}$, where a peak is expected in the PBH Thermal Model \citep{Carr:2019kxo}, and where no black holes from confirmed astrophysical origin have been previously observed. The merger rates associated with these BBH events in the mass-gap have large uncertainties, but can be roughly estimated with the late binary formation in dense clusters of PBH through dynamical captures, and are seen to be in agreement with observations \citep{Clesse:2020ghq}.

In cases like GW230529 where the sky localization is very poorly constrained, systematic monitoring with ground-based optical telescopes is unfeasible. Therefore, even if this system could power a potentially detectable kilonova emission, the lack of complete coverage of the GW sky region prevents us from inferring reliable limits on the brightness of this event in the optical/IR band. 

On the other hand, this study demonstrates how crucial is the role of wide field $\gamma-$ray/X--ray space telescopes, like \textit{Swift} and \textit{Fermi}, for the systematic coverage of the full sky at the moment of a GW merger. The exclusion of a high-energy EM signal can pose stringent constraints on the capability of NSBH mergers to successfully launch a relativistic jet. Moreover, since the GW sources detected by LVK are relatively close, we are potentially sensitive also to the off-axis $\gamma-$ray/X--ray emission, enabling us also to shed more light on the structure of the relativistic outflows launched by these events. In the future, targeted searches performed jointly between \textit{Swift} and \textit{Fermi} will be of primary importance to detect the high-energy EM counterpart of GW mergers, and in the case of non-detection, pose the deepest limits on their intrinsic EM brightness in the $\gamma-$ray/hard X--ray domain.

\section*{Acknowledgments}
The authors are grateful for computational resources provided by the LIGO Laboratory and supported by National Science Foundation Grants PHY-0757058 and PHY-0823459. This material is based upon work supported by NSF's LIGO Laboratory which is a major facility fully funded by the National Science Foundation. J.A.K. and J.D. acknowledge the support of NASA contract NAS5-0136. J.A.K., G.R. and S.R. acknowledge the support of NASA grant 80NSSC24K0488 through the Swift Guest Investigator program. J.A.K., G.R. and S.R. acknowledge the support of NASA grant 80NSSC24K0292 through the Fermi Guest Investigator program. The USRA coauthors gratefully acknowledge NASA funding through contract 80NSSC24M0035. The NASA authors gratefully acknowledge NASA funding through the Fermi-GBM project. The UAH coauthors gratefully acknowledge NASA funding from co-operative agreement 80MSFC22M0004. G.P. acknowledges the partial support by ICSC – Centro Nazionale di Ricerca in High Performance Computing, Big Data and Quantum Computing, funded by European Union – NextGenerationEU. K.C., I.G. R.K. and B.S. were supported by NFS awards AST$\_$2307147, PHY$\_$2308886, PHY$\_$2309064 and PHY$\_$2207638.

\newpage

\appendix

\section{Note about the apparent luminosity structure}
\label{lum_app}
\begin{figure*}
    \centering
    \includegraphics[width=0.48\textwidth]{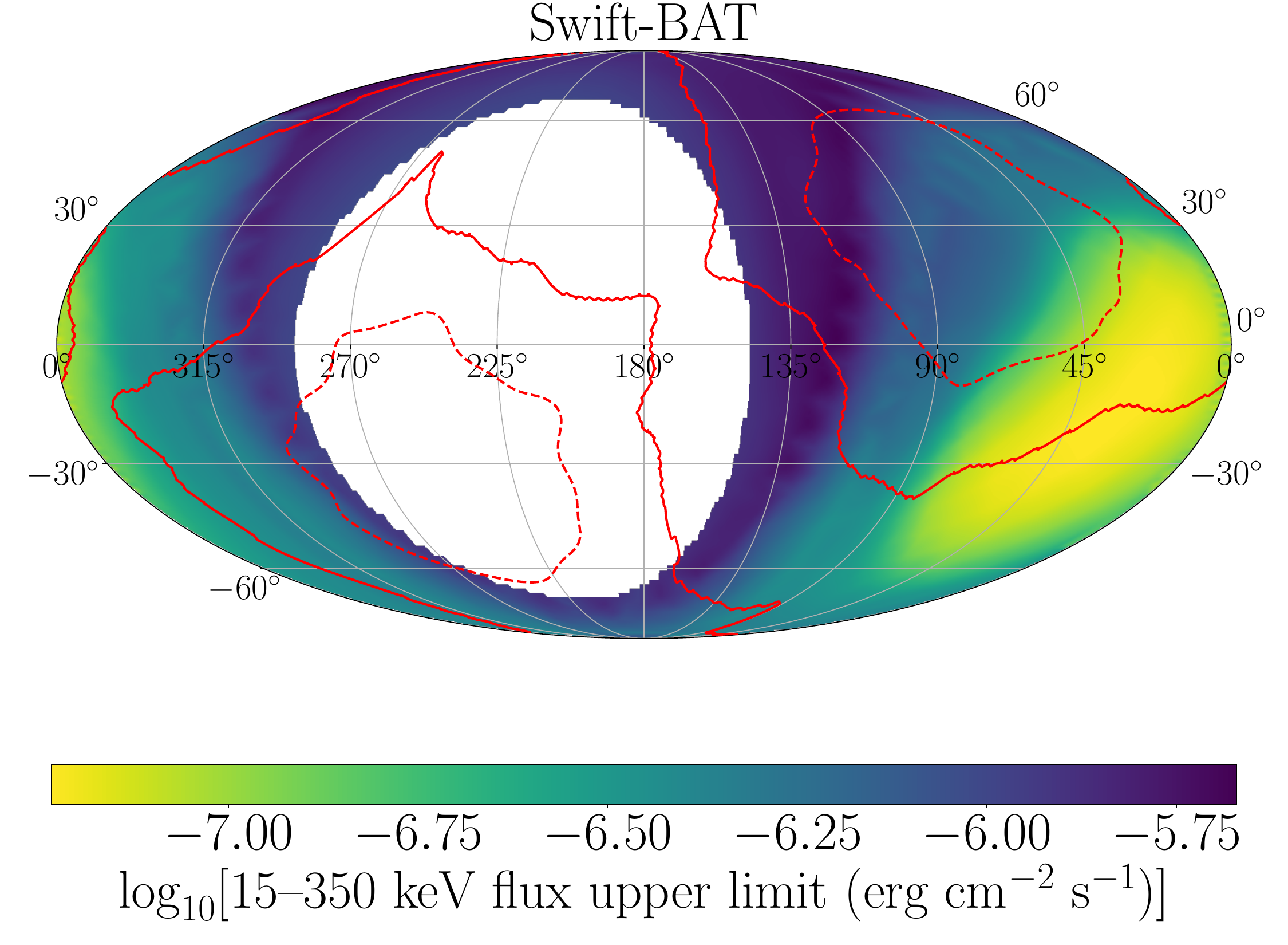} \quad
    \includegraphics[width=0.48\textwidth]{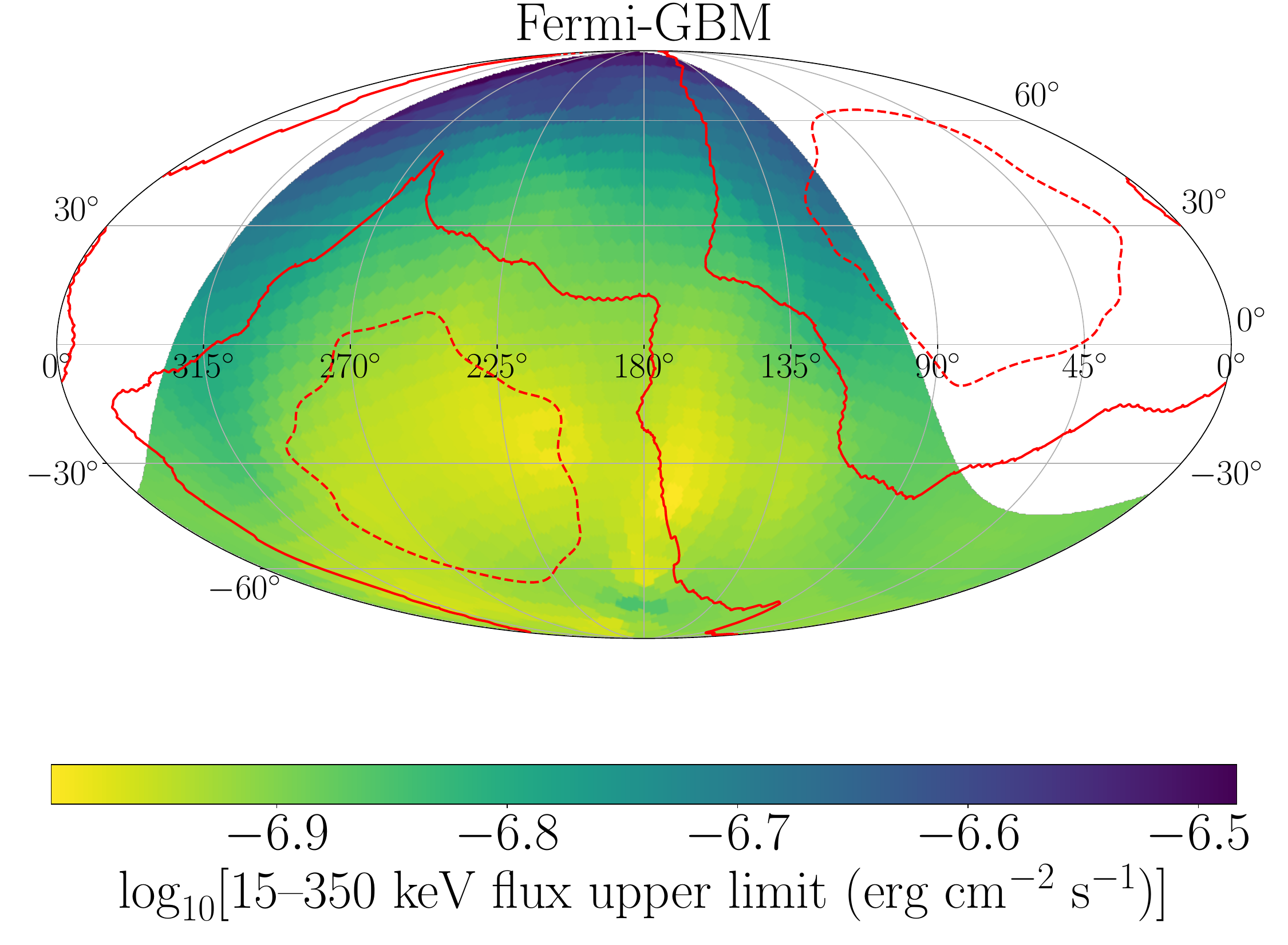}
    \caption{Upper limit sky map in the 15--350 keV band for \textit{Swift}--BAT (left) and \textit{Fermi}--GBM (right). The upper limits are relative to a time scale of 1 second, computed at 5$\sigma$ confidence level. Solid and dashed lines are the GW contours at 90$\%$ and 50$\%$ credibility, respectively. The white region is the position of the Earth, for which no upper limits are available.  }
    \label{fig:ul}
\end{figure*}

Given a structure of the angular distribution of the energy released by the jet during the prompt emission, one can derive the observed flux received by the observer, as a function of the viewing angle. Also the typical duration of the burst $\delta t (\theta_v)$ depends on the viewing angle, therefore one can define the angular structure of the luminosity per unit frequency as
\begin{equation}
    L_{\nu}(\theta_v) = 4 \pi D_L^2 \left < F_{\nu}(\theta_v) \right >_{\delta t (\theta_v)}
\end{equation}
Going from the flux density to the integrated flux in the 15-350 keV band, the apparent luminosity structure considered in this work corresponds to
\begin{equation}
    L(\theta_v) = 4 \pi D_L^2 k(D_L)\left < F_{15-350}(\theta_v) \right >_{\delta t (\theta_v)},
\end{equation}
being 
\begin{equation}
     F_{15-350}(\theta_v) = \int_{15 \,\rm keV }^{350 \, \rm keV} F_{\nu} (\theta_v) d \nu
\end{equation}
and $k(D_L)$ the k-correction.
Therefore the angular dependence of $L(\theta_v)$ implicitly is a combination of the angular dependence of the radiated energy per unit solid angle and the angular dependence of the spectrum, which typically softens as we move away from the jet axis.

Since the flux upper limits are derived  taking the maximum value over different spectral shapes, the derived upper limit on the luminosity has the advantage that it is valid for all the spectra considered, since it is the most conservative. This means that we do not have to make ad-hoc assumptions about the angular dependence of the spectral shape. Therefore, the limits on $L(\theta_v)$ have to be interpreted as spectrum-independent. On the other hand, this implies also that, even if we can derive upper bounds on $L(\theta_v)$, these results are still degenerate with respect to the structure of the radiated energy $E(\Omega)$ and the angular dependence of the observed spectrum $S_{\nu} (\theta_v)$. Namely, given a constraint on $L(\theta_v)$, the corresponding constraint on $E(\Omega)$ still depends on $S_{\nu} (\theta_v)$, and viceversa. On the other hand, once a model for $E(\Omega)$ and $S_{\nu} (\theta_v)$ are specified, an apparent luminosity structure $L(\theta_v)$ can be derived and directly compared with the constraints found in this paper.

\bibliography{bib}{}
\bibliographystyle{aasjournal}

\end{document}